\documentclass{article}
\newcommand{\ket}[1]{|{#1}\rangle}

\begin{document}
\rightline{\vbox{\baselineskip=12pt{\hbox{CALT-68-2112}\hbox{QUIC-97-030}
\hbox{quant-ph/9705031}}}}

\bigskip
\centerline{\Large \bf Reliable Quantum Computers}
\bigskip\bigskip
\centerline{\large John Preskill\footnote{\tt
preskill@theory.caltech.edu}}
\medskip
\centerline{\it California Institute of Technology, Pasadena, CA 91125, USA}
\bigskip
\begin{abstract}
The new field of {\it quantum error correction} has developed spectacularly
since its origin less than two years ago.  Encoded quantum information can be
protected from errors that arise due to uncontrolled interactions with the
environment.  Recovery from errors can work effectively even if occasional
mistakes occur during the recovery procedure.  Furthermore, encoded quantum
information can be {\it processed} without serious propagation of errors.
Hence, an arbitrarily long quantum computation can be performed reliably,
provided that the average probability of error per quantum gate is less than a
certain critical value, the {\it accuracy threshold}. A quantum computer
storing about $10^6$ qubits, with a probability of error per quantum gate of
order $10^{-6}$, would be a formidable factoring engine.  Even a smaller, less
accurate quantum computer would be able to perform many useful tasks.

\smallskip
This paper is based on a talk presented  at the ITP Conference on Quantum
Coherence and Decoherence, 15-18 December 1996.
\end{abstract}

\parskip=5pt 
\section{The golden age of quantum error correction}

Many of us are hopeful that quantum computers will become practical and useful
computing devices some time during the 21st century.  It is probably fair to
say, though, that none of us can now envision exactly what the hardware of that
machine of the future will be like; surely, it will be much different than the
sort of hardware that experimental physicists are investigating these days. But
of one thing we can be quite confident---that a practical quantum computer will
incorporate some type of error correction into its operation.  Quantum
computers are far more susceptible to making errors than conventional digital
computers, and some method of controlling and correcting those errors will be
needed to prevent a quantum computer from crashing.

As recently as mid 1995, we did not have a clear idea how quantum error
correction would work, or whether it would work.  Indeed, there were a number
of reasons for pessimism about whether quantum error correction could really be
possible (Unruh 1995; Landauer 1995, 1996, 1997).  First of all, although very
sophisticated methods have been developed to correct errors in classical
information (MacWilliams \& Sloane 1977), it is not immediately clear how to
adapt these methods to correct the {\it phase} errors that plague quantum
systems.  Second, in a quantum computer, as in a classical analog computer,
small errors can accumulate over time and eventually add up to large errors,
and it is difficult to find methods that can prevent or correct such small
errors.  Third, to correct an error, we must first acquire some information
about the nature of the error by making a measurement, and there is a danger
that the measurement will destroy the delicate quantum information that is
encoded in the device.  Finally, to protect against errors we must encode
information in a redundant manner.  But a famous theorem (Wootters \& Zurek
1982; Dieks 1982) says that quantum information cannot be copied, so it is not obvious how
to store information with the required redundancy.

But by now all of these apparent obstacles have been overcome---we now know
that quantum error correction really is possible.  The key conceptual point we
have grasped is that we can {\it fight entanglement with entanglement}.
Entanglement can be our enemy, since entanglement of our device with the
environment can conceal quantum information from us, and so cause errors.  But
entanglement can also be our friend---we can encode the information that we
want to protect in entanglement, that is, in correlations involving a large
number of qubits.  This information, then, cannot be accessed if we measure
just a few qubits.  By the same token, though, the information cannot be {\it
damaged} if the environment interacts with just a few qubits.  Furthermore, we
have learned that, although the quantum computer is in a sense an analog
device, we can {\it digitalize} the errors that it makes.  We deal with small
errors by making appropriate measurements that project the state of our quantum
computer onto either a state where no error has occurred, or a state with a
large error, which can then be corrected with familiar methods.  And we have
seen that it is possible to measure the errors without measuring the data---we
can acquire information about the precise nature of the error without acquiring
any information about the quantum information encoded in our device (which
would result in decoherence and failure of our computation).  The central idea
of quantum error correction is this:  a small subspace of the Hilbert space of
our device is designated as the {\it code subspace}.  This space is carefully
chosen so that all of the errors that we want to correct move the code space to
mutually orthogonal {\it error subspaces}.  We can make a measurement after our
system has interacted with the environment that tells us in which of these
mutually orthogonal subspaces the system resides, and hence infer exactly what
type of error occurred.  The error can then be repaired by applying an
appropriate unitary transformation.

If we were pessimistic in 1995, there was good reason for optimism by late
1996.  This past year has been a landmark year for quantum information; it is
the year that we have learned how to resist and reverse the effects of
decoherence.  This discovery has important consequences for quantum
computation, but it will also have broader ramifications.  Here are some the
milestones that have been reached this year:  That quantum error correcting
codes exist was first pointed out by Peter Shor (1995) and Andrew Steane
(1996a) in the fall of '95.  By early '96, Steane (1996b) and Calderbank and
Shor (1996) had shown that {\it good} codes exist, that  is, codes that are
capable of correcting many errors.  We learned from the work on random codes by
Lloyd (1997) and by Bennett, DiVincenzo, Smolin, and Wootters (1996) that if we
want to store quantum information for a while, then we can find a code that
will enable us to recover the stored information with high fidelity if the
probability of error per qubit is less than about 19\%.  

But that 19\% estimate of the allowable error rate is quite misleading, because
if applies only under a very unrealistic assumption---that we can encode the
information and perform the recovery {\it flawlessly}, without making any
mistakes.  In fact, encoding and recovery are themselves complex quantum
computations, and errors will inevitably occur while we carry out these
operations.  Thus, we need to find methods for recovering from errors that are
sufficiently robust that we can still recover the quantum information with high
accuracy even when we make some errors during the recovery step.  This is the
problem of {\it fault-tolerant recovery}, and Peter Shor (1996) showed in a
pioneering paper written last May that fault-tolerant recovery is possible if
the error rate is not too high.\footnote{Methods for fault-tolerant recovery were also
developed independently by Alesha Kitaev (1996a).}  Of course, we want more than just to store
quantum information; we want to be able to process the information and build up
an interesting quantum computation.  So we must show that it is possible to
devise quantum gates that work efficiently on information that has been
carefully encoded so as to be protected from errors.  And in the same paper
last May, Shor showed that fault-tolerant {\it computation} is indeed
possible.  

In August, Manny Knill and Raymond Laflamme (1996) showed that there is an
accuracy threshold for storage of quantum information; that is, if the error
rate is below a certain critical value, then it is possible to store an unknown
quantum state with high fidelity for an indefinitely long time.  In the Caltech
group (Gottesman {\it et al.} 1996), we quickly recognized that it is possible
to combine the ideas of Shor with those of Knill and Laflamme to show that
there is also an accuracy threshold for computation; if the error rate is below
a critical value, then it is possible to do an arbitrarily long quantum
computation with a negligible probability of error.  Similar conclusions were
reported by Knill, Laflamme, and Zurek (1996, 1997) , by Aharonov and Ben-Or
(1996), and by Kitaev (1996b).

Hence we are now in a position to sharpen the challenge to designers and
builders of the quantum computers of the future.  We can say just how well the
hardware of that machine will have to work if it is to be used to perform
interesting computations, where we might define interesting as competitive with
what the best digital computers can do today.  But first let's review the basic
elements of fault-tolerant quantum computation.

\section{The laws of fault-tolerant computation}

To perform fault-tolerant computation, we must (1) ensure that recovery from
errors is performed {\it reliably}, (2) be able to implement gates that can
{\it process} encoded information, and (3) control propagation of errors. When
we apply a gate to, say, a pair of qubits, and one of the qubits has an error,
then the error will tend to spread to the other qubit.  We need to be careful
to contain the infection.  The procedures that must be followed to implement
fault-tolerant computation can be codified in what I will call ``the laws of
fault-tolerant computation.'' These can be distilled from Shor's pioneering
paper (Shor 1996).

\begin{figure}
\centering
\begin{picture}(360,80)

\put(0,36){\makebox(20,12){$x$}}
\put(20,40){\line(1,0){40}}
\put(40,40){\circle{8}}
\put(40,44){\line(0,-1){8}}
\put(70,36){\makebox(20,12){$x\oplus 1$}}

\put(120,46){\makebox(20,12){$x$}}
\put(120,26){\makebox(20,12){$y$}}
\put(140,50){\line(1,0){40}}
\put(140,30){\line(1,0){40}}

\put(160,50){\circle*{4}}
\put(160,50){\line(0,-1){24}}
\put(160,30){\circle{8}}

\put(190,26){\makebox(20,12){$x\oplus y$}}
\put(190,46){\makebox(20,12){$x$}}

\put(240,56){\makebox(20,12){$x$}}
\put(240,36){\makebox(20,12){$y$}}
\put(240,16){\makebox(20,12){$x$}}
\put(260,60){\line(1,0){40}}
\put(260,40){\line(1,0){40}}
\put(260,20){\line(1,0){40}}

\put(280,60){\circle*{4}}
\put(280,40){\circle*{4}}
\put(280,20){\circle{8}}
\put(280,60){\line(0,-1){44}}

\put(310,56){\makebox(20,12){$x$}}
\put(310,36){\makebox(20,12){$y$}}
\put(310,16){\makebox(20,12){$z\oplus xy$}}

\end{picture}
\caption{Diagrammatic notation for the NOT gate, the XOR (controlled-NOT) gate,
and the Toffoli (controlled-controlled-NOT) gate.}
\label{fig_gates}
\end{figure}
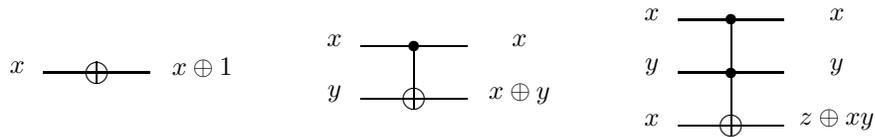

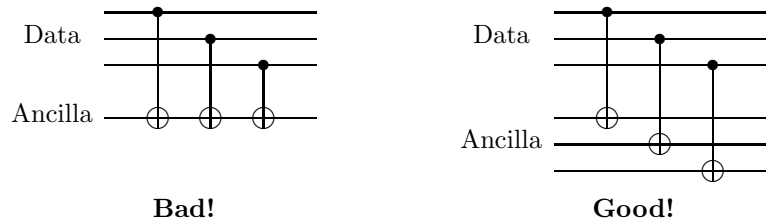
\begin{figure}
\centering
\begin{picture}(280,85)

\put(50,0){\makebox(20,12){\bf Bad!}}
\put(0,36){\makebox(20,12){Ancilla}}
\put(0,66){\makebox(20,12){Data}}

\put(30,40){\line(1,0){80}}
\put(30,60){\line(1,0){80}}
\put(30,70){\line(1,0){80}}
\put(30,80){\line(1,0){80}}

\put(50,80){\circle*{4}}
\put(50,80){\line(0,-1){44}}
\put(50,40){\circle{8}}

\put(70,70){\circle*{4}}
\put(70,70){\line(0,-1){34}}
\put(70,40){\circle{8}}

\put(90,60){\circle*{4}}
\put(90,60){\line(0,-1){24}}
\put(90,40){\circle{8}}

\put(220,0){\makebox(20,12){\bf Good!}}
\put(170,26){\makebox(20,12){Ancilla}}
\put(170,66){\makebox(20,12){Data}}

\put(200,20){\line(1,0){80}}
\put(200,30){\line(1,0){80}}
\put(200,40){\line(1,0){80}}
\put(200,60){\line(1,0){80}}
\put(200,70){\line(1,0){80}}
\put(200,80){\line(1,0){80}}

\put(220,80){\circle*{4}}
\put(220,80){\line(0,-1){44}}
\put(220,40){\circle{8}}

\put(240,70){\circle*{4}}
\put(240,70){\line(0,-1){44}}
\put(240,30){\circle{8}}

\put(260,60){\circle*{4}}
\put(260,60){\line(0,-1){44}}
\put(260,20){\circle{8}}

\end{picture}
\caption{The first law. A bad circuit that uses the same ancilla bit several
times, and a good circuit that uses each ancilla bit only once.}
\label{first}
\end{figure}

The first law is {\bf (1) Don't use the same bit twice}.\footnote{A less snappy but more judicious version of this law would be:  Avoid using the same qubit too many times.}  A bad network of XOR
gates that breaks this commandment is shown in Fig.~\ref{first} (using the
notation of Fig.~\ref{fig_gates}).  The bit that I have called the ancilla is
the target of several successive gates.  If there is a single error in this
ancilla bit, this network may propagate that error to the several other bits
that are the sources of the XOR gates; hence the infection spreads virulently.
A peculiarly quantum-mechanical feature is that, while even a classical XOR
gate will propagate bit flip errors from the source to the target, for quantum
gates we must also worry about phase errors, and the phase errors propagate in
the opposite direction, from the target to the source.  So in quantum
computation we need to be especially careful about propagation of errors.  To
follow this law, we may redesign the network as shown, expanding the ancilla to
several bits so that no single bit is acted on more than once.

\begin{figure}
\centering
\begin{picture}(240,80)

\put(0,16){\makebox(20,12){Ancilla}}
\put(0,46){\makebox(20,12){Data}}

\put(30,20){\line(1,0){20}}
\put(30,50){\line(1,0){80}}

\put(50,5){\framebox(40,30){}}

\put(70,25){\makebox(0,0){Prepare}}
\put(70,15){\makebox(0,0){State}}

\put(90,20){\line(1,0){20}}

\put(110,5){\framebox(70,60){}}

\put(145,50){\makebox(0,0){Error}}
\put(145,35){\makebox(0,0){Syndrome}}
\put(145,20){\makebox(0,0){Computation}}

\put(180,20){\line(1,0){20}}
\put(180,50){\line(1,0){20}}

\put(210,16){\makebox(20,12){Measure}}

\end{picture}
\caption{The second law. The ancilla must be prepared properly, so that the
fault-tolerant syndrome measurement diagnoses the errors without spoiling the
coherence of the data.}
\label{second}
\end{figure}
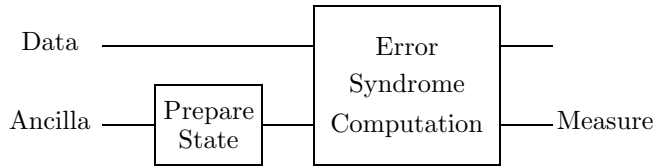

The second law concerns how recovery from errors must be effected.  To recover,
we copy some information from the data to an ancilla and then measure the
ancilla to find an {\it error syndrome} that tells us what recovery operation
is required.  The second law says that we should {\bf (2) Copy the errors, not
the data}.  If some of the encoded information is copied to the ancilla, the
resulting entanglement of the data register and ancilla will cause decoherence
and hence errors.  To avoid this, we must prepare a special state of the
ancilla before we copy any information, chosen so that by measuring the ancilla
we acquire only information about the errors that have occurred, and learn
nothing about the encoded data. (See Fig.~\ref{second}.)

\begin{figure}
\centering
\begin{picture}(320,100)

\put(0,26){\makebox(20,12){$|0\rangle$}}
\put(0,46){\makebox(20,12){$|0\rangle$}}
\put(0,66){\makebox(20,12){$|0\rangle$}}

\put(20,30){\line(1,0){20}}
\put(20,50){\line(1,0){20}}
\put(20,70){\line(1,0){20}}

\put(40,25){\framebox(40,50){}}

\put(60,50){\makebox(0,0){Encode}}

{\thicklines
\put(80,50){\line(1,0){50}}
}

\put(130,25){\framebox(40,50){}}
\put(150,60){\makebox(0,0){Verify}}
\put(150,40){\makebox(0,0){$|0\rangle_{\rm code}$}}
\put(125,30){\line(1,0){5}}
\put(115,30){\makebox(0,0){Anc.}}

\put(170,30){\line(1,0){5}}
\put(175,30){\line(0,-1){15}}
\put(175,5){\makebox(0,0){Meas.}}

{\thicklines
\put(170,50){\line(1,0){50}}
}

\put(220,25){\framebox(40,50){}}
\put(240,60){\makebox(0,0){Verify}}
\put(240,40){\makebox(0,0){$|0\rangle_{\rm code}$}}

\put(215,30){\line(1,0){5}}
\put(205,30){\makebox(0,0){Anc.}}

\put(260,30){\line(1,0){5}}
\put(265,30){\line(0,-1){15}}
\put(265,5){\makebox(0,0){Meas.}}

{\thicklines
\put(260,50){\line(1,0){20}}
}
\put(300,50){\makebox(0,0){$|0\rangle_{\rm code}$}}

\end{picture}
\caption{The third law. The encoder constructs $|0\rangle_{\rm code}$, and then
a nondestructive measurement is performed (at least twice) to verify that the
encoding was successful.}
\label{third}
\end{figure}
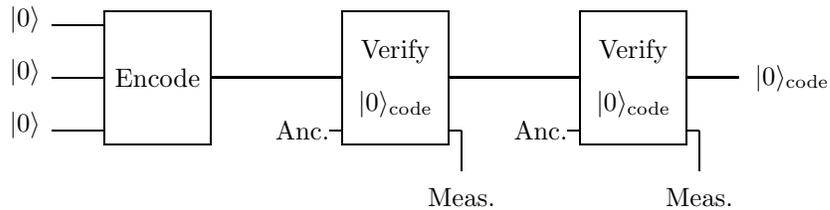

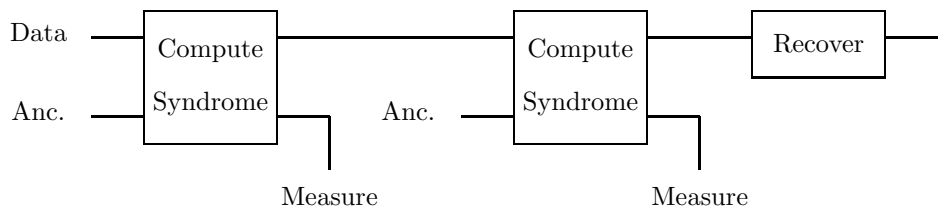
\begin{figure}
\centering
\begin{picture}(350,75)

\put(0,31){\makebox(20,12){Anc.}}
\put(0,61){\makebox(20,12){Data}}

\put(30,35){\line(1,0){20}}
\put(30,65){\line(1,0){20}}

\put(50,25){\framebox(50,50){}}

\put(75,60){\makebox(0,0){Compute}}
\put(75,40){\makebox(0,0){Syndrome}}
\put(100,35){\line(1,0){20}}
\put(120,35){\line(0,-1){20}}
\put(120,5){\makebox(0,0){Measure}}

\put(140,31){\makebox(20,12){Anc.}}

\put(170,35){\line(1,0){20}}
\put(100,65){\line(1,0){90}}

\put(190,25){\framebox(50,50){}}

\put(215,60){\makebox(0,0){Compute}}
\put(215,40){\makebox(0,0){Syndrome}}
\put(240,35){\line(1,0){20}}
\put(260,35){\line(0,-1){20}}
\put(260,5){\makebox(0,0){Measure}}

\put(240,65){\line(1,0){40}}

\put(280,50){\framebox(50,25){}}

\put(330,65){\line(1,0){20}}

\put(305,63){\makebox(0,0){Recover}}

\end{picture}
\caption{The fourth law. Operations such as syndrome measurement must be
repeated to ensure accuracy.}
\label{fourth}
\end{figure}

The third and fourth laws say that whenever we do anything, we should check (if
possible) to make sure it has been done right.  Often we will need to prepare a
block that encodes a known quantum state, such as an encoded 0.  The third law
says that it pays to {\bf (3) {\it Verify} when you encode a {\it known}
quantum state}.  During the encoding procedure we are quite vulnerable to
errors---the power of the code to protect against errors is not yet in place,
and a single error may propagate catastrophically.  Therefore, we should carry
out a measurement that checks that the encoding has been done correctly.  Of
course, the verification itself may be erroneous, so we must repeat the
verification a few times before we have sufficient confidence that the encoding
was correct.

That verification must be repeated is actually a special case of the fourth
law, which says to {\bf (4)  {\it Repeat} operations}.  An important
application of this law is to the measurement of the syndrome that precedes
recovery.  An error during syndrome measurement can both damage the data {\it
and} result in an erroneous syndrome.  If we mistakenly accept the measured
syndrome and act accordingly, we will cause further damage instead of
correcting the error.  Therefore, it is important to be highly confident that
the syndrome measurement is correct before we perform the recovery.  To achieve
sufficient confidence, we must repeat the syndrome measurement several times,
in accord with the fourth law.

\begin{figure}
\centering
\begin{picture}(210,85)

\put(0,26){\makebox(20,12){Data}}
\put(0,66){\makebox(20,12){Data}}

\put(30,20){\line(1,0){80}}
\put(30,30){\line(1,0){80}}
\put(30,40){\line(1,0){80}}
\put(30,60){\line(1,0){80}}
\put(30,70){\line(1,0){80}}
\put(30,80){\line(1,0){80}}

\put(50,80){\circle*{4}}
\put(50,80){\line(0,-1){44}}
\put(50,40){\circle{8}}

\put(70,70){\circle*{4}}
\put(70,70){\line(0,-1){44}}
\put(70,30){\circle{8}}

\put(90,60){\circle*{4}}
\put(90,60){\line(0,-1){44}}
\put(90,20){\circle{8}}

\put(120,46){\makebox(20,12){\Large\bf =}}

{\thicklines
\put(150,30){\line(1,0){60}}
\put(150,70){\line(1,0){60}}

\put(180,70){\circle*{6}}
\put(180,70){\line(0,-1){46}}
\put(180,30){\circle{10}}
}

\end{picture}
\caption{The fifth law. Transversal (bitwise) implementation of a
fault-tolerant XOR gate.}
\label{fifth}
\end{figure}
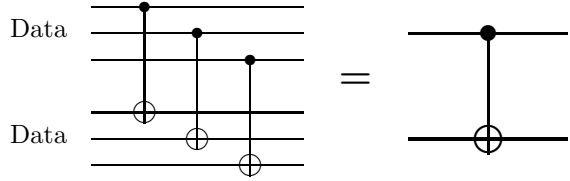

The fifth law is perhaps the most important and also the hardest to obey:  {\bf
(5)  Use the {\it right} code}.  The code that we use for computation should
have special properties so that we can apply quantum gates to the encoded
information that operate efficiently and that adhere to the first four laws.
For example, a good code for computation might be such that an XOR gate acting
on encoded qubits is implemented as in Fig.~\ref{fifth}---with a single XOR
gate applied to each bit in both the source block and the target block.  Then
the gate acting on the encoded qubits obeys the first law and is
fault-tolerant.\footnote{When I gave the talk in December, it was not known how
to implement fault-tolerant gates for most quantum codes.  Since then Gottesman
(1997a) has shown that fault-tolerant computation is possible for all the
``stabilizer codes.''  Still, the fifth law is a good law; any code will do,
but for some codes the fault-tolerant gates are less complicated.} 

\section{\bf Example: Steane's 7-qubit code}

To see how quantum error correction is possible, it is very instructive to
study a particular code. A simple and important example of a quantum
error-correcting code is the 7-qubit code devised by Andrew Steane (1996ab).
This code enables us to store one qubit of quantum information (an arbitrary
state in a two-dimensional Hilbert space) using altogether 7-qubits (by
embedding the two-dimensional Hilbert space in a space of dimension $2^7$).
Steane's code is actually closely related to a familiar classical
error-correcting code, the [7,4,3] Hamming code (MacWilliams \& Sloane 1977).
To understand why Steane's code works, it is important to first understand the
classical Hamming code.

The Hamming code uses a block of 7 bits to encode 4 bits of classical
information; that is, there are $16=2^4$ strings of length 7 that are the valid
codewords.  The codewords can be characterized by a parity check matrix 
\begin{equation}
\label{ham_matrix}
H=\pmatrix{0&0&0&1&1&1&1\cr
	0&1&1&0&0&1&1\cr
	1&0&1&0&1&0&1\cr} \ .
\end{equation}
Each valid codeword is a 7-bit string $v_{\rm code}$ that satisfies
\begin{equation}
\sum_k H_{jk}\left(v_{{\rm code}}\right)_k=0 ~({\rm mod ~ 2}) \ ;
\end{equation}
that is, the matrix $H$ annihilates each codeword in mod 2 arithmetic.  Since
$Z_2=\{0,1\}$ is a (finite) field, familiar results of linear algebra apply
here.   $H$ has three linearly independent rows and its kernel is spanned by
four linearly independent column vectors.  The 16 valid codewords are obtained
by taking all possible linear combinations of these four strings, with
coefficients chosen from $\{0,1\}$.

Now suppose that $v_{\rm code}$ is an (unknown) valid codeword, and that a
single (unknown) error occurs: one of the seven bits flips.  We are assigned
the task of determining which bit flipped, so that the error can be corrected.
This trick can be performed by applying the parity check matrix to the string.
Let $e_i$ denote the string with a one in the $i$th place, and zeros elsewhere.
 Then when the $i$th bit flips, $v_{\rm code}$ becomes $v_{\rm code}+e_i$.  If
we apply $H$ to this string we obtain
\begin{equation}
H\left(v_{\rm code}+ e_i\right)= H e_i
\end{equation}
(because $H$ annihilates $v_{\rm code}$), which is just the $i$th column of the
matrix $H$.  Since all of the columns of $H$ are distinct, we can infer $i$; we
have learned where the error occurred, and we can correct the error by flipping
the $i$th bit back.  Thus, we can recover the encoded data unambiguously if
only one bit flips; but if two or more different bits flip, the encoded data
will be damaged.  It is noteworthy that the quantity $H e_i$ reveals the
location of the error without telling us anything about $v_{\rm code}$; that
is, without revealing the encoded information.

Steane's code generalizes this sort of classical error-correcting code to a
{\it quantum} error-correcting code.  The code uses a 7-qubit ``block'' to
encode one qubit of quantum information, that is, we can encode an arbitrary
state in a two-dimensional Hilbert space spanned by two states: the ``logical
zero'' $|0\rangle_{\rm code}$ and the ``logical one'' $|1\rangle_{\rm code}$.
The code is designed to enable us to recover from an arbitrary error occurring
in any of the 7 qubits in the block.

What do we mean by an arbitrary error?  The qubit in question might undergo a
random {\it unitary} transformation, or it might {\it decohere} by becoming
entangled with states of the environment.  Suppose that, if no error occurs,
the qubit ought be in the state $a|0\rangle +b|1\rangle$.  (Of course, this
particular qubit might be entangled with others, so the coefficients $a$ and
$b$ need not be complex numbers; they can be states that are orthogonal to both
$|0\rangle$ and $|1\rangle$, which we assume are unaffected by the error.)  Now
if the qubit is afflicted by an arbitrary error, the resulting state can be
expanded in the form:
\begin{eqnarray}
\label{superop}
a|0\rangle + b|1\rangle\longrightarrow &\left(a|0\rangle +
b|1\rangle\right)&\otimes \quad |A_{\rm no~error}~~~\rangle_{\rm
env}\nonumber\\
+&\left(a|1\rangle + b|0\rangle\right)&\otimes \quad |A_{\rm
bit-flip}~~~~\rangle_{\rm env}\nonumber\\
+&\left(a|0\rangle - b|1\rangle\right)&\otimes \quad |A_{\rm
phase-flip}~\rangle_{\rm env}\nonumber\\
+&\left(a|1\rangle - b|0\rangle\right)&\otimes \quad 
|A_{\rm both~errors}\rangle_{\rm env} \ ,\nonumber\\
\end{eqnarray}
where each $|A\rangle_{\rm env}$ denotes a state of the environment.  We are
making no particular assumption about the orthogonality or normalization of the
$|A\rangle_{\rm env}$ states,\footnote{Though, of course, the combined
evolution of qubit plus environment is required to be unitary.} so
Eq.~\ref{superop} entails no loss of generality.  We conclude that the
evolution of the qubit can be expressed as a linear combination of four
possibilities:  (1) no error occurs, (2) the bit flip $|0\rangle
\leftrightarrow |1\rangle$ occurs, (3) the relative phase of $|0\rangle$ and
$|1\rangle$ flips, (4) both a bit flip and a phase flip occur.

Now it is clear how a quantum error-correcting code should work (Steane 1996b;
Knill \& Laflamme 1997).  By making a suitable measurement, we wish to diagnose
which of these four possibilities actually occurred.  Of course, in general,
the state of the qubit will be a linear combination of these four states, but
the measurement should project the state onto the basis used in
Eq.~\ref{superop}.  We can then proceed to correct the error by applying one of
the four unitary transformations:
\begin{equation} 
(1)~ {\bf 1} \ , \quad (2) ~ X\equiv\pmatrix{0&1\cr 1&0\cr} \ , \quad (3) ~
Z\equiv\pmatrix{1&0\cr 0& -1\cr} \ , \quad (4)~  X\cdot Z \ ,
\end{equation}
(and the measurement outcome will tell us which one to apply).
By applying this transformation, we restore the qubit to its intended value,
and completely disentangle the quantum state of the qubit from the state of the
environment. It is essential, though, that in diagnosing the error, we learn
nothing about the encoded quantum information, for to find out anything about
the coefficients $a$ and $b$ in Eq.~\ref{superop} would necessarily destroy the
coherence of the qubit.

\begin{figure}
\centering
\begin{picture}(320,115)

\put(0,30){\makebox(0,0){$\ket{0}$}}
\put(70,25){\makebox(20,12){Measure}}

\put(10,110){\line(1,0){270}}
\put(10,100){\line(1,0){270}}
\put(10,90){\line(1,0){270}}
\put(10,80){\line(1,0){270}}
\put(10,70){\line(1,0){270}}
\put(10,60){\line(1,0){270}}
\put(10,50){\line(1,0){270}}

\put(10,30){\line(1,0){50}}

\put(20,80){\circle*{4}}
\put(20,80){\line(0,-1){54}}
\put(20,30){\circle{8}}

\put(30,70){\circle*{4}}
\put(30,70){\line(0,-1){44}}
\put(30,30){\circle{8}}

\put(40,60){\circle*{4}}
\put(40,60){\line(0,-1){34}}
\put(40,30){\circle{8}}

\put(50,50){\circle*{4}}
\put(50,50){\line(0,-1){24}}
\put(50,30){\circle{8}}

\put(110,20){\makebox(0,0){$\ket{0}$}}
\put(180,15){\makebox(20,12){Measure}}

\put(120,20){\line(1,0){50}}

\put(130,100){\circle*{4}}
\put(130,100){\line(0,-1){84}}
\put(130,20){\circle{8}}

\put(140,90){\circle*{4}}
\put(140,90){\line(0,-1){74}}
\put(140,20){\circle{8}}

\put(150,60){\circle*{4}}
\put(150,60){\line(0,-1){44}}
\put(150,20){\circle{8}}

\put(160,50){\circle*{4}}
\put(160,50){\line(0,-1){34}}
\put(160,20){\circle{8}}

\put(220,10){\makebox(0,0){$\ket{0}$}}
\put(290,5){\makebox(20,12){Measure}}

\put(230,10){\line(1,0){50}}

\put(240,110){\circle*{4}}
\put(240,110){\line(0,-1){104}}
\put(240,10){\circle{8}}

\put(250,90){\circle*{4}}
\put(250,90){\line(0,-1){84}}
\put(250,10){\circle{8}}

\put(260,70){\circle*{4}}
\put(260,70){\line(0,-1){64}}
\put(260,10){\circle{8}}

\put(270,50){\circle*{4}}
\put(270,50){\line(0,-1){44}}
\put(270,10){\circle{8}}

\end{picture}
\caption{Computation of the bit-flip syndrome for Steane's 7-qubit code.
Repeating the  computation in the rotated basis diagnoses the phase-flip
errors.  To make the procedure fault tolerant, each ancilla qubit must be
replaced by four qubits in a suitable state.}
\label{fig_syndrome}
\end{figure}
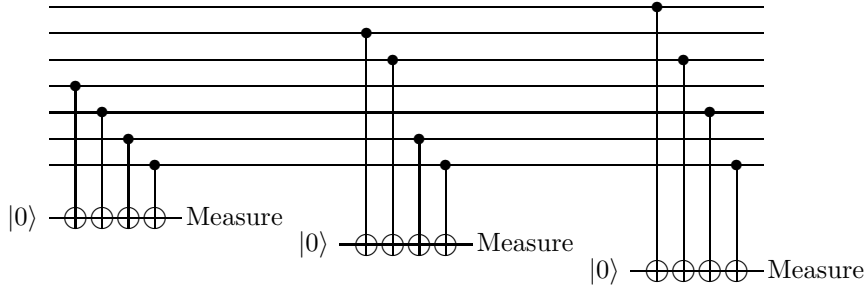

If we use Steane's code, a measurement meeting these criteria is possible.  The
logical zero is the equally weighted superposition of all of the even weight
codewords of the Hamming code (those with an even number of 1's), 
\begin{eqnarray}
\label{zero}
& |0\rangle_{\rm code}={1\over\sqrt{8}} \left(\sum_{{\rm even}~v\atop \in ~{\rm
Hamming}}|v\rangle\right)
={1\over \sqrt{8}} & \Big(  |0000000\rangle + |0001111\rangle +|0110011\rangle
+|0111100\rangle\nonumber\\
 && +  |1010101\rangle  +|1011010\rangle +|1100110\rangle +|1101001\rangle\Big)
\ ,\nonumber\\
\end{eqnarray}
and the logical 1 is the equally weighted superposition of all of the odd
weight codewords of the Hamming code (those with an odd number of 1's),
\begin{eqnarray}
\label{one}
& |1\rangle_{\rm code}={1\over\sqrt{8}}\left(\sum_{{\rm odd}~v\atop \in ~{\rm
Hamming}}|v\rangle \right)
={1\over \sqrt{8}} &\Big(   |1111111\rangle  + |1110000\rangle +|1001100\rangle
+|1000011\rangle\nonumber\\
&& + |0101010\rangle  +|0100101\rangle +|0011001\rangle +|0010110\rangle\Big) \
.\nonumber \\ 
\end{eqnarray}
Since all of the states appearing in Eq.~\ref{zero} and Eq.~\ref{one} are
Hamming codewords, it is easy to detect a single bit flip in the block by doing
a simple quantum computation, as illustrated in Fig.~\ref{fig_syndrome}. We
augment the block of 7 qubits with 3 ancilla bits,\footnote{To make the
procedure fault-tolerant, we will need to increase the number of ancilla bits
as discussed in \S~4.} and perform the unitary operation:
\begin{equation}
|v\rangle\otimes|0\rangle_{\rm anc}\longrightarrow
|v\rangle\otimes|Hv\rangle_{\rm anc} \ ,
\end{equation}
where $H$ is the Hamming parity check matrix, and $|\cdot\rangle_{\rm anc}$
denotes the state of the three ancilla bits. If we assume that only a single
one of the 7 qubits in the block is in error, measuring the ancilla projects
that qubit onto either a state with a bit flip or a state with no flip (rather
than any nontrivial superposition of the two). If the bit does flip, the
measurement outcome diagnoses which bit was affected, without revealing
anything about the quantum information encoded in the block.  

But to perform quantum error correction, we will need to diagnose phase errors
as well as bit flip errors.  To accomplish this, we observe (following Steane
1996ab) that we can change the basis for each qubit by applying the Hadamard
rotation
\begin{equation}
R= {1\over\sqrt{2}}\pmatrix{1&1\cr 1&-1\cr} \ .
\end{equation}
Then phase errors in the $|0\rangle$, $|1\rangle$ basis become bit flip errors
in the rotated basis
\begin{equation}
|\tilde 0\rangle \equiv {1\over\sqrt{2}}\left(|0\rangle +|1\rangle\right) \
,\quad 
|\tilde 1\rangle \equiv {1\over\sqrt{2}}\left(|0\rangle -|1\rangle\right) \ .
\end{equation}
It will therefore be sufficient if our code is able to diagnose bit flip errors
in this rotated basis. But if we apply the Hadamard rotation to each of the 7
qubits, then Steane's logical 0 and logical 1 become in the rotated basis
\begin{eqnarray}
\label{tildezero}
|\tilde 0\rangle_{\rm code}=&{1\over 4}\left(\sum_{v \in \atop {\rm
Hamming}}|v\rangle\right)
={1\over \sqrt{2}}\left(|0\rangle_{\rm code} + |1\rangle_{\rm code}\right) \
,\nonumber\\
|\tilde 1\rangle_{\rm code}=&{1\over 4}\left(\sum_{v \in \atop {\rm
Hamming}}(-1)^{wt(v)}|v\rangle \right)
={1\over \sqrt{2}}\left(|0\rangle_{\rm code} - |1\rangle_{\rm code}\right)\
\nonumber\\
\end{eqnarray}
(where $wt(v)$ denotes the weight of $v$). The key point is that $|\tilde
0\rangle_{\rm code}$ and $|\tilde 1\rangle_{\rm code}$, like $|0\rangle_{\rm
code}$ and $|1\rangle_{\rm code}$, are superpositions of Hamming codewords.
Hence, in the rotated basis, as in the original basis, we can perform the
Hamming parity check to diagnose bit flips, which are phase flips in the
original basis.  Assuming that only one qubit is in error, performing the
parity check in both bases completely diagnoses the error, and enables us to
correct it.

In the above description of the error correction scheme, I assumed that the
error affected only one of the qubits in the block.  Clearly, this assumption
as stated is not realistic; all of the qubits will typically become entangled
with the environment to some degree.   However, as we have seen, the procedure
for determining the error syndrome will typically project each qubit onto a
state in which no error has occurred.  For each qubit, there is a non-zero
probability of an error, assumed small, which we'll call $\epsilon$.  Now we
will make a very important assumption -- that the errors acting on different
qubits in the same block are completely uncorrelated with one another.  Under
this assumption, the probability of two errors is of order $\epsilon^2$, and so
is much smaller than the probability of a single error if $\epsilon$ is small
enough.  So, to order $\epsilon$ accuracy, we can safely confine our attention
to the case where at most one qubit per block is in error.

But in the (unlikely) event of two errors occurring in the same block of the
code, our recovery procedure will typically fail.  If two bits flip in the same
block, then the Hamming parity check will misdiagnose the error.  Recovery will
restore the quantum state to the code subspace, but the {\it encoded}
information in the block will undergo the bit flip
\begin{equation}
|0\rangle_{\rm code} \rightarrow |1\rangle_{\rm code} \ ,
\quad |1\rangle_{\rm code} \rightarrow |0\rangle_{\rm code} \ .
\end{equation}
Similarly, if there are two phase errors in the same block, these are two bit
flip errors in the rotated basis, so that after recovery the block will have
undergone a bit flip in the rotated basis, or in the original basis the phase
flip
\begin{equation}
|0\rangle_{\rm code} \rightarrow |0\rangle_{\rm code} \ ,\quad |1\rangle_{\rm
code} \rightarrow -|1\rangle_{\rm code} \ .
\end{equation}
(If one qubit in the block has a phase error, and another one has a bit flip
error, then recovery will be successful.)

Thus we have seen that Steane's code can enhance the reliability of stored
quantum information.  Suppose that we want to store one qubit in an unknown
pure state $|\psi\rangle$.  Due to imperfections in our storage device, the
state $\rho_{\rm out}$ that we recover will have suffered a loss of fidelity:
\begin{equation}
F\equiv \langle\psi|\rho_{\rm out}|\psi\rangle=1-\epsilon \ .
\end{equation}
But if we store the qubit using Steane's 7-qubit block code, if each of the
7-qubits is maintained with fidelity $F=1-\epsilon$, if the errors on the
qubits are uncorrelated, and if we can perform error recovery, encoding, and
decoding  flawlessly (more on this below), then the encoded information can be
maintained with an improved fidelity $F=1- O\left(\epsilon^2\right)$.  

\begin{figure}
\centering
\begin{picture}(260,140)

\put(25,10){\makebox(0,0){$a \ket{0}+ b \ket{1}$}}
\put(40,30){\makebox(0,0){$\ket{0}$}}
\put(40,50){\makebox(0,0){$\ket{0}$}}
\put(40,70){\makebox(0,0){$\ket{0}$}}
\put(40,90){\makebox(0,0){$\ket{0}$}}
\put(40,110){\makebox(0,0){$\ket{0}$}}
\put(40,130){\makebox(0,0){$\ket{0}$}}

\put(50,10){\line(1,0){100}}
\put(50,30){\line(1,0){100}}
\put(50,50){\line(1,0){100}}
\put(50,70){\line(1,0){100}}
\put(50,130){\line(1,0){14}}
\put(50,110){\line(1,0){14}}
\put(50,90){\line(1,0){14}}

\put(70,10){\circle*{4}}
\put(70,10){\line(0,1){44}}
\put(70,30){\circle{8}}
\put(70,50){\circle{8}}

\put(64,124){\framebox(12,12){$R$}}
\put(64,104){\framebox(12,12){$R$}}
\put(64,84){\framebox(12,12){$R$}}

\put(76,130){\line(1,0){74}}
\put(76,110){\line(1,0){74}}
\put(76,90){\line(1,0){74}}

\put(90,130){\circle*{4}}
\put(90,130){\line(0,-1){124}}
\put(90,10){\circle{8}}
\put(90,30){\circle{8}}
\put(90,70){\circle{8}}

\put(110,110){\circle*{4}}
\put(110,110){\line(0,-1){104}}
\put(110,10){\circle{8}}
\put(110,50){\circle{8}}
\put(110,70){\circle{8}}

\put(130,90){\circle*{4}}
\put(130,90){\line(0,-1){64}}
\put(130,30){\circle{8}}
\put(130,50){\circle{8}}
\put(130,70){\circle{8}}


\put(210,100){\makebox(0,0){$a|0\rangle_{\rm code}$}}
\put(210,70){\makebox(0,0){+}}
\put(210,40){\makebox(0,0){$b|1\rangle_{\rm code}$}}

\end{picture}
\caption{An encoding circuit for Steane's 7-qubit code.}
\label{fig_encode}
\end{figure}
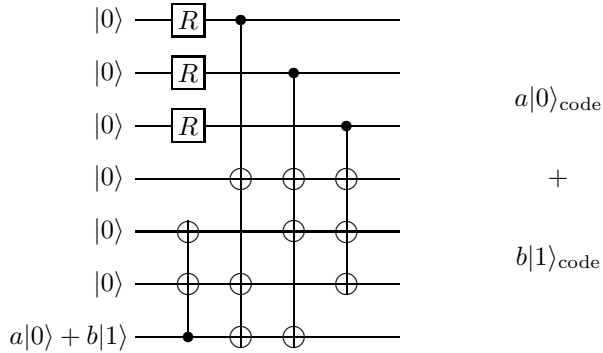

A qubit in an unknown state can be encoded using the circuit shown in
Fig.~\ref{fig_encode}.  It is easiest to understand how the encoder works by
using an alternative expression for the Hamming parity check matrix,
\begin{equation}
H=\pmatrix{1&0&0&1&0&1&1\cr
	0&1&0&1&1&0&1\cr
	0&0&1&1&1&1&0\cr} \ .
\end{equation}
(This form of $H$ is obtained from the form in Eq.~\ref{ham_matrix} by
permuting the columns, which is just a relabeling of the bits in the block.)
The even subcode of the Hamming code is actually the space spanned by the rows
of $H$; so we see that (in this representation of $H$) the first three bits of
the string completely characterize the data represented in the subcode.  The
remaining four bits are the parity bits that provide the redundancy needed to
protect against errors.  When encoding the unknown state $a |0\rangle+
b|1\rangle$, the encoder first uses two XOR's to prepare the state $a
|0000000\rangle+ b|0000111\rangle$, a superposition of even and odd Hamming
codewords. The rest of the circuit adds $|0\rangle_{\rm code}$ to this state:
the Hadamard ($R$) rotations prepare an equally weighted superposition of all
eight possible values for the first three bits in the block, and the remaining
XOR gates switch on the parity bits dictated by $H$. 

We will also want to be able to measure the encoded qubit, say by projecting
onto the orthogonal basis $\{|0\rangle_{\rm code}, |1\rangle_{\rm code}\}$. If
we don't mind destroying the encoded block when we make the measurement, then
it is sufficient to measure each of the seven qubits in the block by projecting
onto the basis $\{|0\rangle,|1\rangle\}$; we then perform classical error
correction on the measurement outcomes to obtain a Hamming codeword.  The
parity of that codeword is the value of the logical qubit.  (The classical
error correction step provides protection against measurement errors. For
example, if the block is in the state $|0\rangle_{\rm code}$, then two
independent errors would have to occur in the measurement of the elementary
qubits for the measurement of the logical qubit to yield the incorrect value
$|1\rangle_{\rm code}$.)

\begin{figure}
\centering
\begin{picture}(360,115)

\put(0,70){\line(1,0){30}}
\put(0,60){\line(1,0){30}}
\put(0,50){\line(1,0){30}}
\put(0,40){\line(1,0){30}}
\put(0,30){\line(1,0){30}}
\put(0,20){\line(1,0){30}}
\put(0,10){\line(1,0){30}}

\put(40,65){\makebox(20,12){Measure}}
\put(40,55){\makebox(20,12){Measure}}
\put(40,45){\makebox(20,12){Measure}}
\put(40,35){\makebox(20,12){Measure}}
\put(40,25){\makebox(20,12){Measure}}
\put(40,15){\makebox(20,12){Measure}}
\put(40,5){\makebox(20,12){Measure}}

\put(70,70){\line(1,0){30}}
\put(70,60){\line(1,0){30}}
\put(70,50){\line(1,0){30}}
\put(70,40){\line(1,0){30}}
\put(70,30){\line(1,0){30}}
\put(70,20){\line(1,0){30}}
\put(70,10){\line(1,0){30}}

\put(100,5){\framebox(40,70){}}
\put(120,50){\makebox(0,0){\it Classical}}
\put(120,30){\makebox(0,0){Recovery}}

\put(140,40){\line(1,0){20}}
\put(165,35){\makebox(20,12){0 or 1}}

\put(240,90){\line(1,0){60}}
\put(240,80){\line(1,0){60}}
\put(240,70){\line(1,0){60}}
\put(240,60){\line(1,0){60}}
\put(240,50){\line(1,0){60}}
\put(240,40){\line(1,0){60}}
\put(240,30){\line(1,0){60}}

\put(250,10){\line(1,0){40}}

\put(240,10){\makebox(0,0){$\ket{0}$}}
\put(300,5){\makebox(20,12){Measure}}

\put(260,50){\circle*{4}}
\put(260,50){\line(0,-1){44}}
\put(260,10){\circle{8}}

\put(270,40){\circle*{4}}
\put(270,40){\line(0,-1){34}}
\put(270,10){\circle{8}}

\put(280,30){\circle*{4}}
\put(280,30){\line(0,-1){24}}
\put(280,10){\circle{8}}

\put(330,80){\makebox(0,0){$|0\rangle_{\rm code}$}}
\put(330,60){\makebox(0,0){or}}
\put(330,40){\makebox(0,0){$|1\rangle_{\rm code}$}}


\end{picture}
\caption{Destructive and nondestructive measurement of the logical qubit.}
\label{fig_measure}
\end{figure}
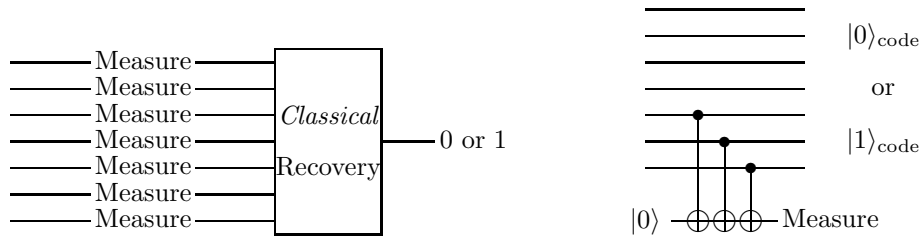

In applications to quantum computation, we will need to perform a measurement
that projects onto $\{|0\rangle_{\rm code}, |1\rangle_{\rm code}\}$ without
destroying the block.  This task is accomplished by copying the parity of the
block onto an ancilla qubit, and then measuring the ancilla. A circuit that
performs a nondestructive measurement of the code block is shown in
Fig.~\ref{fig_measure}.

Steane's 7-qubit code can recover from only a single error in the code block,
but better codes can be constructed that can protect the information from up to
$t$ errors within a single block, so that the encoded information can be
maintained with a fidelity $F=1 - O\left(\epsilon^{t+1}\right)$ (Steane 1996b;
Calderbank and Shor 1996; Gottesman 1996; Calderbank {\it et al.} 1996, 1997).
The current status of quantum coding theory is reviewed by Shor in these
proceedings (Shor 1997).

\section{Fault-tolerant recovery}
\label{sec_recovery}

But, of course, error recovery will never be flawless.  Recovery is itself a
quantum computation that will be prone to error.  We must take care to design a
recovery procedure for Steane's code that ensures that the probability of
failure is of order $\epsilon^2$ (or of order $\epsilon^{t+1}$ for a code that
corrects $t$ errors). Most critically, we must control propagation of error
during the recovery procedure. 

The circuit shown in Fig.~\ref{fig_syndrome} is not fault tolerant; it violates
the first law.  The XOR gates can propagate a single phase error in one of the
ancilla bits to two different qubits in the data block, resulting in a phase
error in the block that can occur with probability of order $\epsilon$.  To
adhere to the first law, we must expand each ancilla bit to four distinct bits,
each of which is the target of just a single XOR gate.  But now we must respect
the second law: we must entangle the ancilla with the {\it errors} in the data
block, but not with the quantum information encoded there, for entanglement of
the ancilla with the encoded data will destroy the coherence of the data.  

To meet this criterion, we prepare a {\it Shor state} of the ancilla before
proceeding with the syndrome computation (Shor 1996). This is a state of four
ancilla bits that is an equally weighted superposition of all even weight
strings:
\begin{equation}
|{\rm Shor}\rangle_{\rm anc}={1\over\sqrt{8}}\sum_{{\rm even}~v}|v\rangle_{\rm
anc} \ .
\end{equation}
To compute each bit of the syndrome, we prepare a Shor state, perform four XOR
gates (with appropriate qubits in the data block as the sources and the four
bits of the Shor state as the targets), and then measure the ancilla state.
The syndrome bit is inferred from the parity of the four measured ancilla bits.
 The Shor state has been chosen so that {\it only} this parity can be inferred
from the state of the ancilla---no other information about the data block gets
imprinted there.

\begin{figure}
\centering
\begin{picture}(210,85)

\put(10,30){\line(1,0){19}}
\put(29,24){\framebox(12,12){$R$}}
\put(41,30){\line(1,0){38}}

\put(10,70){\line(1,0){19}}
\put(29,64){\framebox(12,12){$R$}}
\put(41,70){\line(1,0){38}}

\put(79,24){\framebox(12,12){$R$}}
\put(91,30){\line(1,0){19}}

\put(79,64){\framebox(12,12){$R$}}
\put(91,70){\line(1,0){19}}

\put(60,70){\circle*{6}}
\put(60,70){\line(0,-1){45}}
\put(60,30){\circle{10}}

\put(120,46){\makebox(20,12){\Large\bf =}}

\put(150,30){\line(1,0){60}}
\put(150,70){\line(1,0){60}}
\put(180,30){\circle*{6}}
\put(180,30){\line(0,1){45}}
\put(180,70){\circle{10}}

\end{picture}
\caption{A useful identity. The source and the target of an XOR gate are interchanged if we perform a change of basis with Hadamard rotations.}
\label{identity}
\end{figure}
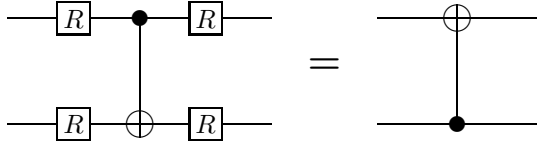

There are altogether 6 syndrome bits (3 to diagnose bit-flip errors and 3 to
diagnose phase-flip errors), so the syndrome measurement uses 24 ancilla bits
prepared in 6 Shor states, and 24 XOR gates. 

(One way to obtain the phase-flip syndrome would be to first apply 7 parallel $R$ gates to the data block to rotate the basis, then to apply the XOR gates as in Fig.~\ref{fig_syndrome} (but with the ancilla expanded into a Shor state), and finally to apply 7 $R$ gates to rotate the data back.  However, we can use the identity represented in Fig.~\ref{identity} to improve this procedure.  By reversing the direction of the XOR gates (that is, by using the ancilla as the source and the data as the target), we can avoid applying the $R$ gates to the data, and hence can reduce the likelihood of damaging the data with faulty gates (Zalka 1996; Steane 1997).) 

\begin{figure}
\centering
\begin{picture}(210,120)

\put(0,14){\makebox(20,12){$\ket{0}$}}
\put(0,34){\makebox(20,12){$\ket{0}$}}
\put(0,54){\makebox(20,12){$\ket{0}$}}
\put(0,74){\makebox(20,12){$\ket{0}$}}
\put(0,94){\makebox(20,12){$\ket{0}$}}

\put(20,20){\line(1,0){140}}
\put(20,40){\line(1,0){134}}
\put(20,60){\line(1,0){134}}
\put(20,80){\line(1,0){134}}
\put(20,100){\line(1,0){14}}
\put(46,100){\line(1,0){108}}

\put(34,94){\framebox(12,12){$R$}}

\put(60,100){\circle*{4}}
\put(60,100){\line(0,-1){24}}
\put(60,80){\circle{8}}

\put(80,80){\circle*{4}}
\put(80,80){\line(0,-1){24}}
\put(80,60){\circle{8}}

\put(100,60){\circle*{4}}
\put(100,60){\line(0,-1){24}}
\put(100,40){\circle{8}}

\put(120,100){\circle*{4}}
\put(120,100){\line(0,-1){84}}
\put(120,20){\circle{8}}

\put(140,40){\circle*{4}}
\put(140,40){\line(0,-1){24}}
\put(140,20){\circle{8}}

\put(154,34){\framebox(12,12){$R$}}
\put(154,54){\framebox(12,12){$R$}}
\put(154,74){\framebox(12,12){$R$}}
\put(154,94){\framebox(12,12){$R$}}
\put(160,14){\makebox(40,12){Measure}}

\put(166,40){\line(1,0){14}}
\put(166,60){\line(1,0){14}}
\put(166,80){\line(1,0){14}}
\put(166,100){\line(1,0){14}}

\end{picture}
\caption{Construction and verification of the Shor state.  If the measurement
outcome is 1, then the state is discarded and a new Shor state is prepared.}
\label{fig-4qubitcat}
\end{figure}
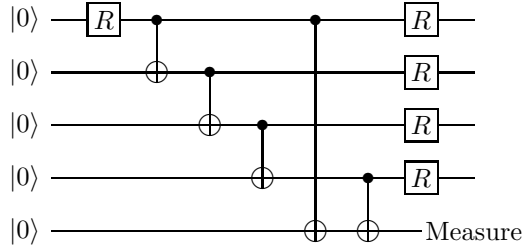

Due to error propagation, a single error that occurs during the preparation of
the Shor state can result in two phase errors in the state, and these can both
propagate to the data if the faulty ancilla is used for syndrome measurement.
Therefore the Shor state must be tested for multiple phase errors before it is
used (an instance of the third law), as in Fig.~\ref{fig-4qubitcat}. If it
fails the test, it should be discarded, and a new Shor state should be
constructed.

Finally, a single error during syndrome measurement can result in a faulty
syndrome.  Thus, the syndrome measurement should be repeated to verify accuracy
(the fourth law).  When the same result is obtained twice in a row, the result
can be safely accepted and recovery can proceed.

If we take all the precautions described above, then recovery will fail only if
two independent errors occur, so the probability of an error occurring in the
encoded block will be of order $\epsilon^2$.  The error-correction procedure is
fault-tolerant.

A somewhat streamlined fault-tolerant syndrome measurement procedure was
suggested by Steane (1997).  For the measurement of the bit-flip syndrome, we
prepare the ancilla in the 7-qubit {\it Steane state}
\begin{equation}
|{\rm Steane}\rangle_{\rm anc}={1\over 4}\sum_{v\in ~{\rm Hamming}} |v\rangle \
.
\end{equation}
(The Steane state may be prepared by applying the bitwise Hadamard rotation to
the state $|0\rangle_{\rm code}$.)
We may then XOR each qubit of the data block into the corresponding qubit of
the ancilla, and measure the ancilla. Applying the Hamming parity-check matrix
$H$ to the {\it classical} measurement outcome, we obtain the bit-flip
syndrome. (Note the adherence to the second law: the procedure ``copies'' the
data onto the ancilla, but the state of the ancilla has been carefully chosen
to ensure that only the information about the error can be read by measuring
the ancilla.) The same procedure is carried out in the rotated basis to find
the phase-flip syndrome.  The Steane procedure has the advantage over the Shor
procedure that only 14 ancilla bits and 14 XOR gates are needed.  But it also
has the disadvantage that the ancilla preparation is more complex, so that the
ancilla is somewhat more prone to error.

What about measurement and encoding?  We have already noted that destructive
measurement of the code block will be reliable if only one qubit in the block
is in error.  The nondestructive measurement depicted in Fig.~\ref{fig_measure}
also need not be modified.  Though the ancilla is the target of three
successive XOR gates, phase errors feeding back into the block are not harmful
because they cannot change $|0\rangle_{\rm code}$ to $|1\rangle_{\rm code}$ (or
vice versa).  However, since a single error can cause a faulty parity
measurement, the measurement must be repeated (after error correction) to
ensure accuracy to order $\epsilon^2$ (an instance of the fourth law).

In applications to quantum computation, we will need to repeatedly prepare the
encoded state $|0\rangle_{\rm code}$.  The encoding may be performed with the
circuit in Fig.~\ref{fig_encode} (except that the first two XOR gates may be
eliminated).  However errors can propagate during the encoding procedure, so
that a single error may suffice to cause an error in the encoded data.  Hence
it is important to verify the encoding, as stated in the third law, by
performing nondestructive measurement of the block.  In fact, the encoding step
can actually be dispensed with.  Whatever the initial state of the block,
fault-tolerant error correction will project it onto the space spanned by
$\{|0\rangle_{\rm code},|1\rangle_{\rm code}\}$, and (verified) measurement
will project out either $|0\rangle_{\rm code}$ or $|1\rangle_{\rm code}$.  If
the result $|1\rangle_{\rm code}$ is obtained, then the (bitwise) NOT operator
can be applied to flip the block to the desired state $|0\rangle_{\rm code}$.

If we wish to encode an unknown quantum state, then we use the encoding circuit
in Fig.~\ref{fig_encode}.  Again, because of error propagation, a single error
during encoding may cause an encoding failure.  In this case, since no
measurement can verify the encoding, the fidelity of the encoded state will
inevitably be $F=1-O(\epsilon)$.  However, encoding may still be worthwhile,
since it may enable us to preserve the state with a reasonable fidelity for a
longer time than if the state had remained unencoded.

Both Shor's and Steane's scheme for fault-tolerant syndrome measurement have
been described here only for the 7-qubit code, but they can be adapted to more
complex codes that have the capability to recover from many errors (DiVincenzo
\& Shor 1996; Steane 1997).  As the complexity of the code increases, Steane's scheme becomes substantially more efficient than Shor's.

\section{Fault-tolerant quantum gates}
\label{sec:gates}

We have seen that coding can protect quantum information. But we want to do
more than {\it store} quantum information with high fidelity; we want to
operate a quantum computer that {\it processes} the information.   Of course,
we could decode, perform a gate, and then re-encode, but that procedure would
temporarily expose the quantum information to harm.  Instead, if we want our
quantum computer to operate reliably, we must be able to apply quantum gates
directly to the encoded data, and these gates must respect the first law of
fault tolerance if catastrophic propagation of error is to be avoided.

In fact, with Steane's 7-qubit code, there are a number of gates that can be
easily implemented.  Three single-qubit gates can all be applied {\it bitwise};
that is applying these gates to each of the 7 qubits in the block implements
the same gate acting on the encoded qubit.  We have already seen in
Eq.~\ref{tildezero} that the Hadamard rotation $R$ acts this way.  The same is
true for the NOT gate (since each odd parity Hamming codeword is the complement
of an even parity Hamming codeword)\footnote{Actually, we can implement the NOT
acting on the encoded qubit with just 3 NOT's applied to selected qubits in the
block.}, and the phase gate
\begin{equation}
P=\pmatrix{1&0\cr 0& i\cr} \ ;
\end{equation}
(the odd Hamming codewords have weight $\equiv$ 3 (mod 4) and the even
codewords have weight $\equiv$ 0 (mod 4), so we actually apply $P^{-1}$ bitwise
to implement $P$).  The XOR gate can also be implemented bitwise; that is, by
XOR'ing each bit of the source block into the corresponding bit of the target
block.  This works because the even codewords form a subcode, while the odd
codewords are its nontrivial coset.

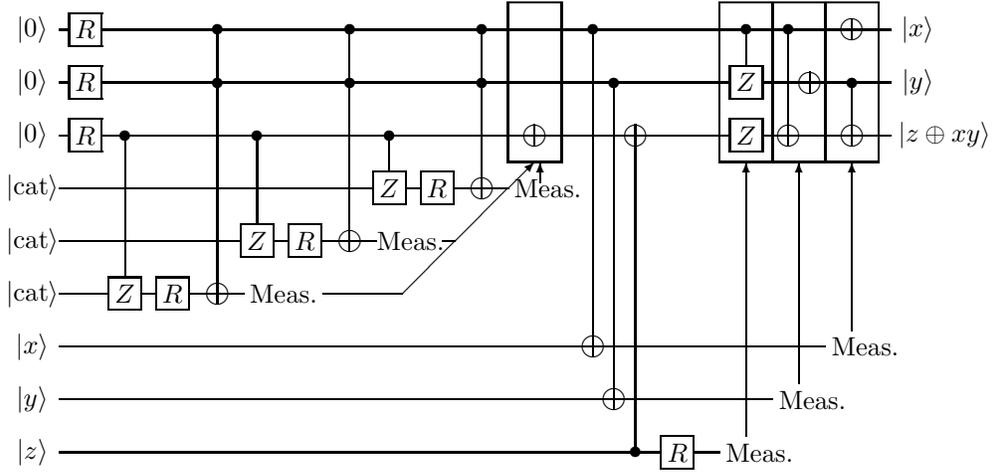
\begin{figure}
\centering
\begin{picture}(335,200)

\put(0,174){\makebox(20,12){$\ket{0}$}}
\put(0,154){\makebox(20,12){$\ket{0}$}}
\put(0,134){\makebox(20,12){$\ket{0}$}}
\put(0,114){\makebox(20,12){$\ket{\rm cat}$}}
\put(0,94){\makebox(20,12){$\ket{\rm cat}$}}
\put(0,74){\makebox(20,12){$\ket{\rm cat}$}}
\put(0,54){\makebox(20,12){$\ket{x}$}}
\put(0,34){\makebox(20,12){$\ket{y}$}}
\put(0,14){\makebox(20,12){$\ket{z}$}}

\put(20,180){\line(1,0){4}}
\put(24,174){\framebox(12,12){$R$}}
\put(20,160){\line(1,0){4}}
\put(24,154){\framebox(12,12){$R$}}
\put(20,140){\line(1,0){4}}
\put(24,134){\framebox(12,12){$R$}}

\put(36,180){\line(1,0){299}}
\put(36,160){\line(1,0){238}}
\put(286,160){\line(1,0){49}}
\put(36,140){\line(1,0){238}}
\put(286,140){\line(1,0){49}}

\put(20,80){\line(1,0){19}}
\put(51,80){\line(1,0){6}}
\put(69,80){\line(1,0){21}}
\put(57,74){\framebox(12,12){$R$}}

\put(39,74){\framebox(12,12){$Z$}}
\put(45,140){\circle*{4}}
\put(45,140){\line(0,-1){54}}

\put(80,180){\circle*{4}}
\put(80,180){\line(0,-1){104}}
\put(80,160){\circle*{4}}
\put(80,80){\circle{8}}

\put(90,74){\makebox(30,12){Meas.}}

\put(20,100){\line(1,0){69}}
\put(101,100){\line(1,0){6}}
\put(119,100){\line(1,0){21}}
\put(107,94){\framebox(12,12){$R$}}

\put(89,94){\framebox(12,12){$Z$}}
\put(95,140){\circle*{4}}
\put(95,140){\line(0,-1){34}}

\put(130,180){\circle*{4}}
\put(130,180){\line(0,-1){84}}
\put(130,160){\circle*{4}}
\put(130,100){\circle{8}}

\put(138,94){\makebox(30,12){Meas.}}

\put(20,120){\line(1,0){119}}
\put(151,120){\line(1,0){6}}
\put(169,120){\line(1,0){21}}
\put(157,114){\framebox(12,12){$R$}}

\put(139,114){\framebox(12,12){$Z$}}
\put(145,140){\circle*{4}}
\put(145,140){\line(0,-1){14}}

\put(180,180){\circle*{4}}
\put(180,180){\line(0,-1){64}}
\put(180,160){\circle*{4}}
\put(180,120){\circle{8}}

\put(190,114){\makebox(30,12){Meas.}}

\put(200,140){\circle{8}}
\put(200,136){\line(0,1){8}}
\put(190,130){\framebox(20,60){}}
\put(150,80){\vector(1,1){50}}
\put(120,80){\line(1,0){30}}
\put(165,100){\line(1,0){5}}
\put(202,122){\vector(0,1){8}}

\put(20,60){\line(1,0){290}}
\put(20,40){\line(1,0){270}}
\put(20,20){\line(1,0){228}}

\put(222,180){\circle*{4}}
\put(222,180){\line(0,-1){124}}
\put(222,60){\circle{8}}

\put(230,160){\circle*{4}}
\put(230,160){\line(0,-1){124}}
\put(230,40){\circle{8}}

\put(238,20){\circle*{4}}
\put(238,20){\line(0,1){124}}
\put(238,140){\circle{8}}

\put(248,14){\framebox(12,12){$R$}}
\put(260,20){\line(1,0){10}}

\put(270,14){\makebox(30,12){Meas.}}
\put(290,34){\makebox(30,12){Meas.}}
\put(310,54){\makebox(30,12){Meas.}}

\put(280,180){\circle*{4}}
\put(280,180){\line(0,-1){14}}
\put(274,154){\framebox(12,12){$Z$}}
\put(274,134){\framebox(12,12){$Z$}}
\put(270,130){\framebox(20,60){}}
\put(280,26){\vector(0,1){104}}

\put(296,180){\circle*{4}}
\put(296,180){\line(0,-1){44}}
\put(296,140){\circle{8}}
\put(304,160){\circle{8}}
\put(304,156){\line(0,1){8}}
\put(290,130){\framebox(20,60){}}
\put(300,46){\vector(0,1){84}}

\put(320,180){\circle{8}}
\put(320,176){\line(0,1){8}}
\put(320,160){\circle*{4}}
\put(320,160){\line(0,-1){24}}
\put(320,140){\circle{8}}
\put(310,130){\framebox(20,60){}}
\put(320,66){\vector(0,1){64}}

\put(335,174){\makebox(20,12){$\ket{x}$}}
\put(335,154){\makebox(20,12){$\ket{y}$}}
\put(335,134){\makebox(40,12){$\ket{z \oplus xy}$}}

\end{picture}
\caption[The fault-tolerant Toffoli gate.]{The fault-tolerant Toffoli gate.
Each 
line represents a block of 7 qubits, and the gates are implemented
transversally. For each measurement, the arrow points to the set of gates that
is to be applied if the measurement outcome is 1; no action is taken if the
outcome is 0.}
\label{fig-toffoli}
\end{figure}

Thus there are simple fault-tolerant procedures for implementing the gates NOT,
$R$, $P$, and XOR.  But unfortunately, these gates do not by themselves form a
universal set.  To be able to perform arbitrary unitary transformations on
encoded data, we will need to make a suitable addition to this set.  Following,
Shor (1996), we will add the 3-qubit Toffoli gate, which is implemented by the
procedure shown in Fig.~\ref{fig-toffoli}.\footnote{Knill {\it et al.} (1996,
1997) describe an alternative way of completing the universal set of gates.}  

Briefly, the procedure works as follows.  First, three encoded ancilla blocks
are prepared in a state of the form
\begin{equation}
|A\rangle_{\rm anc}\equiv \sum_{a=0,1}\sum_{b=0,1} |a,b,ab\rangle_{\rm anc} \
.
\end{equation}
This preparation of the ancilla is performed by using a (verified) 7-bit ``cat
state'' 
\begin{equation}
|{\rm cat}\rangle={1\over\sqrt{2}}\bigl(|0000000\rangle + |1111111\rangle\bigr)
\ .
\end{equation}
A few gates are performed, including a bitwise Toffoli gate with two ancilla
blocks as the controls and the cat state as the target.  Then the cat state is
measured.  (The measurement is repeated to ensure accuracy.) If the parity of
the measurement outcomes is even, then the desired ancilla state
$|A\rangle_{\rm anc}$ has been successfully prepared; if the parity is odd,
then the state $|A\rangle_{\rm anc}$ can be obtained by applying a NOT to the
third ancilla block.

Meanwhile, three data blocks have been waiting patiently for the ancilla to be
ready.  By applying three XOR gates and a Hadamard rotation, the state of the
data and ancilla is transformed as 
\begin{equation}
\sum_{a=0,1}\sum_{b=0,1} |a,b,ab\rangle_{\rm anc}|x,y,z\rangle_{\rm
data}\longrightarrow
\sum_{a=0,1}\sum_{b=0,1}\sum_{w=0,1} (-1)^{wz} |a,b,ab\oplus z\rangle_{\rm
anc}|x\oplus a,y\oplus b,w\rangle_{\rm data} \ .
\end{equation}
Now each {\it data} block is measured.  If the measurement outcome is 0, no
action is taken, but if the measurement outcome is 1, then a particular set of
gates is applied to the {\it ancilla}, as shown in Fig.~\ref{fig-toffoli}, to
complete the implementation of the Toffoli gate.  Note that the original data
blocks are destroyed by the procedure, and that what were initially the ancilla
blocks become the new data blocks. The important thing about this construction
is that all of the steps have been carefully designed to adhere to the laws of
fault tolerance.

That the fault-tolerant gates form a discrete set is a bit of a nuisance, but
it is also an unavoidable feature of any fault-tolerant scheme.  It would not
make sense for the fault-tolerant gates to form a continuum, for then how could
we possibly avoid making an error by applying the {\it wrong} gate, a gate that
differs from the intended one by a small amount?  Anyway, since our
fault-tolerant gates form a universal set,  they suffice for approximating any
desired unitary transformation to any desired accuracy.

Shor described how to generalize this fault tolerant set of gates to more
complex codes that can correct more errors, and  Gottesman (1997ab) has
described an even more general procedure that can be applied to any of the
known quantum codes.  Thus, almost any quantum error-correcting code can be
used for fault-tolerant computation.  What then is the meaning of the fifth
law?  While any code can be used in principle, some codes are better than
others.  For example, there is a 5-qubit code that can recover from one error
(Bennett {\it et al.} 1996, Laflamme {\it et al.} 1996), and Gottesman has
exhibited a universal set of fault-tolerant gates for this code.  But the gate
implementation is quite complex.  The 7-qubit Steane code requires a larger
block, but it is much more convenient for computation; the fifth law dictates
that the Steane code should be preferred.

\section{The accuracy threshold for quantum computation}

Quantum error-correcting codes exist that can correct $t$ errors, where $t$ can
be arbitrarily large.  If we use such a code and we follow the laws of
fault-tolerance, then an uncorrectable error will occur only if at least $t+1$
{\it independent} errors occur in a single block before recovery is completed.
So if the probability of an error occurring per quantum gate, or the
probability of a storage error occurring per unit of time, is of order
$\epsilon$, then the probability of an error per gate acting on encoded data
will be of order $\epsilon^{t+1}$, which is much smaller if $\epsilon$ is small
enough.  Indeed, it may seem that by choosing a code with $t$ as large as we
please we can make the probability of error per gate as small as we please, but
this turns out not to be the case, at least not for most codes.  The trouble is
that as we increase $t$, the complexity of the code increases sharply, and the
complexity of the recovery procedure correspondingly increases.  Eventually we
reach the point where it takes so long to perform recovery that it is likely
that $t+1$ errors will accumulate in a block before we can complete the
recovery step, and the ability of the code to correct errors is thus
compromised.

Suppose that the number of computational steps needed to perform the syndrome
measurement increases with $t$ like a power $t^b$.  Then the probability that
$t+1$ errors accumulate before the measurement is complete will behave like
\begin{equation}
\label{t_prob}
{\rm Block ~Error ~Probability}\sim \left( t^b \epsilon\right)^{t+1}\  ,
\end{equation}
where $\epsilon$ is the probability of error per step.
We may then choose $t$ to minimize the error probability ($t\sim
e^{-1}\epsilon^{-1/b}$, assuming $t$ is large), obtaining
\begin{equation}
{\rm Minimum ~Block ~Error ~Probability}\sim
\exp\left(-e^{-1}b\epsilon^{-1/b}\right) \ .
\end{equation}
Thus if we hope to carry out altogether $T$ cycles of error correction without
any error occurring, 
then our gates must have an accuracy
\begin{equation}
\label{acc_need}
\epsilon\sim \left(\log T\right)^{-b} \ .
\end{equation}
Similarly, if we hope to perform a quantum computation with altogether $T$
quantum gates, elementary gates of this prescribed accuracy are needed.

In the procedure originally described by Shor (1996), the power characterizing
the complexity of the syndrome measurement is $b=4$, and modest improvements
($b\sim 3$) can be achieved with a  better optimized procedure.   The block
size of the code used grows with $t$ like $t^2$, so when the code is chosen to
optimize the error probability, the block size is of order $(\log T)^2$.
Certainly, the scaling described by  Eq.~\ref{acc_need} is much more favorable
than the accuracy $\epsilon\sim T^{-1}$ that would be required were coding not
used at all.  But for any given accuracy, there is a limit to how long a
computation can proceed until errors become likely.

\begin{figure}
\centering
\begin{picture}(360,230)

\put(0,230){\line(1,0){80}}
\put(0,210){\line(1,0){80}}
\put(0,190){\line(1,0){23}}
\put(57,190){\line(1,0){23}}

{\thicklines
\put(40,190){\circle{30}}
}
\put(26,192){\line(1,0){28}}
\put(26,190){\line(1,0){28}}
\put(26,188){\line(1,0){28}}

{\thicklines
\put(40,170){\line(0,-1){20}}
\put(50,150){\oval(20,20)[bl]}
\put(50,140){\vector(1,0){60}}
}

\put(120,160){\line(1,0){80}}
\put(120,140){\line(1,0){80}}
\put(120,120){\line(1,0){23}}
\put(177,120){\line(1,0){23}}

{\thicklines
\put(160,120){\circle{30}}
}

\put(146,122){\line(1,0){28}}
\put(146,120){\line(1,0){28}}
\put(146,118){\line(1,0){28}}

{\thicklines
\put(160,100){\line(0,-1){20}}
\put(170,80){\oval(20,20)[bl]}
\put(170,70){\vector(1,0){60}}
}

\put(240,90){\line(1,0){80}}
\put(240,70){\line(1,0){80}}
\put(240,50){\line(1,0){23}}
\put(297,50){\line(1,0){23}}

{\thicklines
\put(280,50){\circle{30}}
}

\put(266,48){\line(1,0){28}}
\put(266,50){\line(1,0){28}}
\put(266,52){\line(1,0){28}}

{\thicklines
\put(280,30){\line(0,-1){20}}
\put(290,10){\oval(20,20)[bl]}
\put(290,0){\vector(1,0){60}}
}

\end{picture}
\caption{Concatenated coding.  Each qubit in the block, when inspected at
higher resolution, is itself an encoded subblock.}
\label{fig_concatenate}
\end{figure}
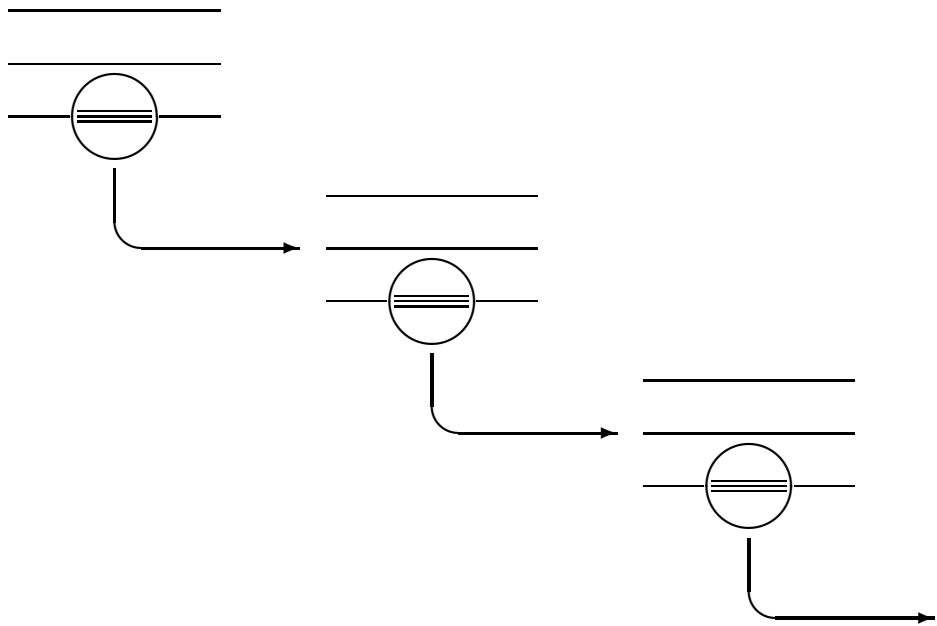

This limitation can be overcome by using a special kind of code, a {\it
concatenated} code (Knill \& Laflamme 1996; Knill {\it et al.} 1996, 1997;
Aharonov \& Ben-Or 1996; Kitaev 1996ab).   To understand the concept of a
concatenated code, imagine that we are using Steane's quantum error-correcting
code that encodes a single qubit as a block of $7$ qubits.  But if we look more
closely at one of the $7$ qubits in the block with better resolution, we
discover that it is not really a single qubit, but another block of $7$ encoded
using the same Steane code as before.  And when we examine one of the $7$
qubits in {\it this} block with higher resolution, we discover that it too is
really a block of $7$ qubits.  And so on. (See Fig.~\ref{fig_concatenate}.) If
there are all together $L$ levels to this hierarchy of concatenation, then a
single qubit is actually encoded in a block of size $7^L$.  The reason that
concatenation is useful is that it enables us to recover from errors more
efficiently, by dividing and conquering.  That is, we perform recovery most
often at the lowest level of the hierarchy, on a block of size $7$, less often
at the next highest level, on a block of size $7^2=49$, still less often at the
next level, on a block of size $7^3=343$, and so on.  With this method, the
complexity of error correction does not grow so sharply as we increase the
error-correcting capacity of our quantum code.

We have seen that Steane's 7-qubit code can recover from one error.  If the
probability of error per qubit is $\epsilon$, the errors are uncorrelated, and
recovery is fault-tolerant, then the probability of a recovery failure is of
order $\epsilon^2$.  If we concatenate the code to construct a block of size
$7^2$, then an error occurs in the block only if two of the subblocks of size 7
fail, which occurs with a probability of order $(\epsilon^2)^2$.  And if we
concatenate again, then an error occurs only if two subblocks of size $7^2$
fail, which occurs with a probability of order $((\epsilon^2)^2)^2$.  If there
are all together $L$ levels of concatenation, then the probability of an error
is or order $(\epsilon)^{2^L}$, while the block size is $7^L$.  Now, if the
error rate for our fundamental gates is small enough, then we can improve the
probability of an error per gate by concatenating the code.  If so, we can
improve the performance even more by adding another level of concatenation.
And so on.    This is the origin of the accuracy threshold for  quantum
computation:  if coding reduces the probability of error significantly, then we
can make the error rate arbitrarily small by adding enough levels of
concatenation.  But if the error rates are too high to begin with, then coding
will make things worse instead of better.

To analyze this situation, we must first adopt a particular model of the
errors,  and I will choose the simplest possible quasi-realistic model:
uncorrelated stochastic errors.\footnote{I will comment in \S~\ref{dream} on
how the analysis is modified if a different error model is adopted.}  In  each
computational time step, each qubit in the device becomes entangled with the
environment as in Eq.~\ref{superop}, but where we assume that the four states
of the environment are mutually orthogonal, and that the three ``error states''
have equal norms.  Thus the three types of errors (bit flip, phase flip, both)
are assumed to be equally likely.  The total probability of error in each time
step is denoted $\epsilon_{\rm store}$.  In addition to these storage errors
that afflict the ``resting'' qubits, there will also be errors that are
introduced  by the quantum gates themselves.  
For each type of gate, the probability of error each time the gate is
implemented is denoted $\epsilon_{\rm gate}$ (with independent values assigned
to gates of each type).  If the gate acts on more than one qubit (XOR or
Toffoli), correlated errors may arise. In our analysis, we make the pessimistic assumption 
that an error in a multi-qubit gate always damages all of the qubits on which the gate acts;
{\it e.g.}, a faulty XOR gate introduces errors in both the source qubit and the target qubit. This assumption (among others) is made only to keep the analysis tractable.  Under more realistic assumptions, we would surely find that somewhat higher error rates could be tolerated.

We can analyze the efficacy of concatenated coding by constructing a set of
{\it concatenation flow equations}, that describe how our error model evolves
as we proceed from one level of concatenation to the next.  For example,
suppose we want to perform an XOR gate followed by an error recovery step on
qubits encoded using the concatenated Steane code with $L$ levels of
concatenation (block size $7^L$).   The gate implementation and recovery can be
described in terms of operations that act on subblocks of size $7^{L-1}$.
Thus, we can derive an expression for the probability of error $\epsilon^{(L)}$
for a gate acting on the full block in terms of the probability of error for
gates acting on the subblocks.  This expression is one of the flow equations.
In principle, we can solve the flow equations to find the error probabilities
``at level $L$'' in terms of the parameters of the error model for the
elementary qubits. We then study how the error probabilities behave as $L$
becomes large.  If all block error probabilities approach zero for $L$ large,
then the elementary error probabilities are ``below the threshold.''  Since our
elementary error model may have many parameters, the threshold is really a
codimension one surface in a high-dimension space.

The flow equations are intricate; for a detailed description see Gottesman
(1997b), Gottesman {\it et al.} (1996), or Zalka (1996).   But to crudely
illustrate the idea, let us suppose that there are no storage errors, and that
the only gates that cause errors are the XOR gates, with a probability of error
per gate given by $\epsilon_{\rm XOR}$.  Suppose that we wish to implement a
network of XOR gates that act on encoded blocks.  Let us estimate how small
$\epsilon_{\rm XOR}$ must be for the error rate achieved by gates acting on
encoded data to be smaller than the error rate for the elementary XOR gates.

Using Shor's ancilla state, 12 XOR gates are needed to compute either the bit
flip or phase flip syndrome,\footnote{This number would be reduced to 7 if
Steane's ancilla state were used; however, since the preparation of the Steane
state is more complex, the ancilla would be more prone to error.} so the
probability of error for each syndrome measurement is
\begin{equation}
\epsilon_{\rm syndrome}\sim 12 \epsilon_{\rm XOR} \ .
\end{equation}
To ensure that the syndrome is reliable, it is measured repeatedly until the
same result is obtained twice in a row.\footnote{This is not really the optimal
procedure, but we will adopt it here for simplicity.} From some simple
combinatorics, one sees that the probability that two independent errors occur
before the syndrome measurement is successfully completed is $5\epsilon_{\rm
syndrome}^2$.  If we assume that both errors damage the data, then two errors
will result in an uncorrectable error in the block.  Since we need to measure
both the bit-flip syndrome and the phase-flip syndrome, we see that the
probability of failure each time we perform recovery is
\begin{equation}
\epsilon_{\rm failure}\sim 2\cdot 5\cdot (12\epsilon_{\rm XOR})^2= 1440
\epsilon_{\rm XOR}^2 \ .
\end{equation}
In the implementation of a network of XOR gates, we will perform recovery on
each block after it has been acted on by $N$ gates, where $N$ is a number that
will be chosen to optimize the error per gate.   Each transversal XOR gate
acting on a block is performed by implementing 7 elementary XOR gates.   When
we perform $N$ gates followed by recovery on a given block, the probability per
gate that two errors accumulate in the block is no worse than 
{\nobreak
\begin{eqnarray}
\label{fail_prob}
\epsilon_{\rm fail} && \sim{1\over N}\left( {1\over 2} 7N\cdot 6N\cdot
\epsilon_{\rm XOR}^2 + 7N\cdot \epsilon_{\rm XOR} \cdot 2 \cdot 2\cdot\epsilon_{\rm
syndrome} + 1440 \cdot\epsilon_{\rm XOR}^2\right)\nonumber\\
&& ={1\over N}\left(21 N^2+336N +
1440\right)\epsilon_{\rm XOR}^2\nonumber\\
\end{eqnarray}}
(There might be two errors introduced by gates, one from a gate and one from
the recovery step, or two during recovery;\footnote{Of the two factors of 2 in the second term of Eq.~\ref{fail_prob}, one arises because we measure both the bit-flip and phase-flip syndromes, the other because the error can occur during either the first or second repetition of the syndrome measurement.} we pessimistically assume that all the errors damage
the data in the block, and that recovery always fails if two errors occur within a data block.)
The minimum of Eq.~\ref{fail_prob} occurs if we choose $N=8$, and we then
obtain $\epsilon_{\rm fail}=684 \epsilon_{\rm XOR}^2$.  So coding pays off if
$684\epsilon_{\rm XOR}^2< \epsilon_{\rm XOR}$, or $\epsilon_{\rm XOR} <
(684)^{-1}\sim 1.5 \cdot 10^{-3}$.  This is our ``threshold estimate.''

Even if all errors {\it were} due to XOR gates, this estimate would be
inadequate for a number of reasons.  Most importantly, we have assumed that the
ancilla states that are used in syndrome measurement are error free, while in
fact XOR gates are used in the preparation and verification of the Shor state.
Furthermore, when XOR gates act on the subblocks of a concatenated code, it is
not always possible to implement the optimal number of gates before recovery.
And of course a more complete analysis would incorporate the storage errors,
and would track the coupled flow of storage and gate errors as a level of
concatenation is added.

Indeed, when we perform recovery at level-$L$, the Shor states must be
constructed from blocks that are encoded at level-$(L-1)$. An important part of
the analysis is to monitor carefully how noisy these ancilla states are, in
order to determine how likely it is for recovery to fail at level-$L$.  The
data blocks need to wait while the ancilla states are prepared; meanwhile
storage errors are accumulating in the data.  Because the waiting period does
not scale simply with the level $L$, the flow equations are not entirely
self-similar --- there is some explicit ``time-dependence'' (that is,
$L$-dependence) in the flow.  It is still roughly true, though, that the
threshold is established by demanding that the fault-tolerant procedure with a
single level of concatenation really improves the reliability of the gates, as
in the above example. 

Though the flow equations are too complicated to solve exactly, we can obtain
approximate solutions by making conservative assumptions. Crudely speaking, the
conclusions are as follows (Gottesman {\it et al.} 1996; Gottesman 1997b): If
storage errors are negligible, then the threshold rate for gate errors is about
$10^{-4}$. If storage errors dominate, then the threshold error rate is about
$10^{-5}$ per time step (where the unit of time is the time required to
implement one gate). The elementary Toffoli gates are not required to be as
accurate as the one and two-body gates -- an Toffoli gate error rate of
$10^{-3}$ is acceptable, if the other error rates are sufficiently small. (This
finding is welcome, since Tofolli gates are more difficult to implement, and
are likely to be less accurate in practice.)

We should also ask how large a block size is needed to ensure a certain
specified accuracy. Roughly speaking, if the threshold gate error rate is
$\epsilon_0$ and the actual elementary gate error rate is $\epsilon<
\epsilon_0$, then concatenating the code $L$ times will reduce the error rate
to
\begin{equation}
\epsilon^{(L)}\sim  \epsilon_0\left({\epsilon\over\epsilon_0}\right)^{2^L} \ ;
\end{equation}
thus, to be reasonably confident that we can complete a computation with $T$
gates without making an error we must choose the block size $7^L$ to be 
\begin{equation}
\label{scaling}
{{\rm block}\atop{\rm size}} \sim \left[ {\log \epsilon_0 T}\over {\log
\epsilon_0/\epsilon}\right]^{\log_2 7} \quad .
\end{equation}
If the code that is concatenated has block size $n$ and can correct $t+1$
errors, the power  $\log_2 7\sim 2.8$ in Eq.~\ref{scaling} is replaced by $\log
n / \log (t+1)$;  this power approaches 2 for the family of codes considered by
Shor, but could in principle approach 1  for ``good'' codes.

When the error rates are below the accuracy threshold, it is also possible to
maintain an {\it unknown} quantum state for an indefinitely long time.
However, as we have already noted in \S~\ref{sec_recovery}, if the probability
of a storage error per computational time step is $\epsilon$, then the initial
encoding of the state can be performed with a fidelity no better than
$F=1-O(\epsilon)$.  With concatenated coding, we can store unknown quantum
information with reasonably good fidelity for an indefinitely long time, but we
cannot achieve arbitrarily good fidelity.

The assumptions that underlie these conclusions will be reviewed in \S
{}~\ref{dream}.

\section{Fault-tolerant factorization}

To get an idea what the scaling law Eq.~\ref{scaling} might mean in practice,
consider using concatenated coding to implement Shor's (1994) quantum factoring
algorithm.  The algorithm has two parts.  To factor the number $N$, we must
first evaluate the modular exponential function to prepare a state of the form
\begin{equation}
\sum_x |x\rangle_{\rm input}\otimes |a^x ~ ({\rm mod} ~ N)\rangle_{\rm output}
\ ,
\label{eq:mod}
\end{equation}
(where $a<N$ is relatively prime to $N$), and then perform a Fourier transform
acting on the input register.  Finally, we measure the input register, and do
some classical post-processing to find a candidate factor of $N$.  The
preparation of the state Eq.~\ref{eq:mod} (using the fault tolerant gates
listed in \S~\ref{sec:gates}) is described in Beckman {\it et al.} (1996) and
Vedral {\it et al.} (1996).  The Fourier transform requires more comment.  As
shown by Griffiths and Niu (1996), the Fourier transform can be evaluated with
single-qubit gates (but with the type of gate conditioned on the outcome of
previous measurements).  In particular, we will need to be able to perform
phase rotation gates of the form
\begin{equation}
P(\theta)=\pmatrix{1 & 0 \cr 0 &e^{i\theta}\cr} \ .
\end{equation}
It suffices to have a fault-tolerant procedure for implementing $P(\theta_0)$
for a particular $\theta_0$ that is an irrational multiple of $2\pi$, since
then applying $P(\theta_0)$ repeatedly allows us to come arbitrarily close to
$P(\theta)$ for any value of $\theta$.

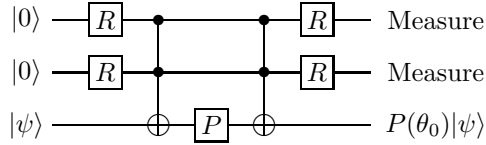
\begin{figure}
\centering
\begin{picture}(180,80)

\put(0,14){\makebox(20,12){$\ket{\psi}$}}
\put(0,34){\makebox(20,12){$\ket{0}$}}
\put(0,54){\makebox(20,12){$\ket{0}$}}

\put(20,60){\line(1,0){14}}
\put(20,40){\line(1,0){14}}
\put(20,20){\line(1,0){54}}

\put(34,34){\framebox(12,12){$R$}}
\put(34,54){\framebox(12,12){$R$}}

\put(46,60){\line(1,0){68}}
\put(46,40){\line(1,0){68}}

\put(60,40){\circle*{4}}
\put(60,60){\circle*{4}}
\put(60,60){\line(0,-1){44}}
\put(60,20){\circle{8}}

\put(100,40){\circle*{4}}
\put(100,60){\circle*{4}}
\put(100,60){\line(0,-1){44}}
\put(100,20){\circle{8}}

\put(74,14){\framebox(12,12){$P$}}
\put(86,20){\line(1,0){54}}

\put(126,60){\line(1,0){14}}
\put(126,40){\line(1,0){14}}

\put(114,34){\framebox(12,12){$R$}}
\put(114,54){\framebox(12,12){$R$}}

\put(145,34){\makebox(40,12){Measure}}
\put(145,54){\makebox(40,12){Measure}}
\put(145,14){\makebox(40,12){$P(\theta_0)|\psi\rangle$}}

\end{picture}
\caption{A rotation circuit.  If the measurement outcome is $(00)$, then the
rotation $P(\theta_0)$ is applied to the data qubit, where
$\cos(\theta_0)$=3/5.}
\label{fig_rotation}
\end{figure}

To construct $P(\theta_0)$ we need to use ancilla bits and apply the Toffoli
gate.  One circuit that works is shown in Fig.~\ref{fig_rotation}.  When the
two ancilla bits are measured, the probability is $5\over 8$ that the outcome
is $|00\rangle_{\rm ancilla}$.  If this is the outcome then the circuit
succeeds in applying $P(\theta_0)$ to the data qubit, where
$e^{i\theta_0}=(1+3i)/(3+i)$, or $\cos(\theta_0)={3\over 5}$.  For any other
outcome ($|01\rangle$, $|10\rangle$, or $|11\rangle$, each occurring with
probability $1\over 8$), the circuit fails.  But we can repair the damage to
the qubit by applying the $Z$-gate, and then make further attempts to apply
$P(\theta_0)$ until we finally succeed.

\begin{figure}
\centering
\begin{picture}(120,50)

\put(0,14){\makebox(20,12){$\ket{\theta}$}}
\put(0,34){\makebox(20,12){$\ket{\psi}$}}

\put(20,40){\line(1,0){40}}
\put(20,20){\line(1,0){40}}

\put(40,40){\circle*{4}}
\put(40,40){\line(0,-1){24}}
\put(40,20){\circle{8}}

\put(65,14){\makebox(40,12){Measure}}
\put(65,34){\makebox(40,12){$P(\theta)|\psi\rangle$}}

\end{picture}
\caption{A rotation circuit that employs an ancilla qubit withdrawn from the
angle library.  If the measurement outcome is 0, then $P(\theta)$ is applied to
the data qubit.}
\label{fig_library}
\end{figure}
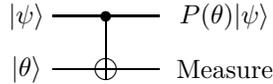

An alternate method is to assemble ``off-line'' a library of ``angle qubits''
of the form 
\begin{equation}
|\theta\rangle={1\over \sqrt{2}}\left( |0\rangle + e^{i\theta}|1\rangle
\right)\ .
\end{equation}
To apply $P(\theta)$ to a data qubit, we perform a controlled-NOT with the data
qubit as the control and the library qubit $|\theta\rangle$ as the target, as
in Fig.~\ref{fig_library}.  We then measure the library qubit.  With
probability ${1\over 2}$, the outcome of the measurement will be $|0\rangle$,
and if so then we have successfully applied $P(\theta)$ to the data qubit. But
if the outcome is $|1\rangle$, then we have applied $P(-\theta)$ instead.  In
the event of failure, we make another attempt, but this time we try to apply
$P(2\theta)$, to compensate for the error in the previous step.  The
probability of failing $n$ times in a row is only $2^{-n}$.

Now let us attempt to factor a $K$-bit number using the factoring circuit of
Beckman {\it et al.} (1996), employing the first method described above to
perform the single-qubit rotations required in the Fourier transform.  To
perform the Fourier transform (on a register with $2K$ qubits), $2K$
single-qubit phase rotations are needed. The Fourier transform need not be
evaluated with perfect precision; for the factoring algorithm to have a
reasonable chance of succeeding, it is sufficient for  each rotation to have a
precision of order $K^{-1}$.  We can achieve this accuracy by composing
$P(\theta_0)$ of order $K$ times.  Hence the Fourier transform requires of
order $K^2$ Toffoli gates, and we conclude that for large $K$, the complexity
of the algorithm is dominated by the evaluation of the modular exponential
function, which requires $38 K^3$ Toffoli gates according to Beckman {\it et
al.} (1996).\footnote{The algorithm described by Beckman {\it et al.} actually requires $46 K^3$ Toffoli gates, but this can be reduced to $38 K^3$ using an improved comparison algortihm suggested by Richard Hughes (1997).}

With the best known classical algorithms and the fastest existing machines, it
takes of order a few months to factor a 130 digit ($K\sim 430$ bit) number
(Lenstra {\it et al.} 1996).  Let's ask what resources a quantum computer would
need to perform this task.  The machine would need to be able to store $5K\sim
2150$ encoded qubits, and to perform of order $3 \cdot 10^9$ Toffoli gates.  For there
to be a reasonable probability of performing the computation without an error,
we would like the probability of error per Toffoli gate to be less than about
$10^{-9}$, and the probability of a storage error per gate execution time to be
less than about $10^{-12}$.  According to the concatenation flow equations,
these error rates can be achieved for the encoded data, if the error rates at
the level of individual qubits are $\epsilon_{\rm store}\sim \epsilon_{\rm
gate}\sim 10^{-6}$, and if 3 levels of concatenation are used, so that the size
of the block encoding each qubit is $7^3=343$.  Allowing for the additional
ancilla qubits needed to implement gates and (parallelized) error correction,
the total number of qubits required in the machine would be of order $10^6$.

With the error rates of order $10^{-6}$ for the individual qubits,
concatenating the 7-qubit code may be the most effective fault-tolerant
procedure.  For error rates about an order of magnitude smaller, we might do
better with a more complex (unconcatenated) code that can correct several
errors in a single block (Shor 1996).  At still lower error rates it is
possible to use codes that make more efficient use of storage space by encoding
many qubits in a single block (Gottesman 1997a).  In fact, when the error rate
for individual qubits is very low, it becomes possible in principle to find a
good code for fault-tolerant computation such that the ratio of encoded qubits
to total qubits approaches one.

\section{Plumbing quantum leaks}
\label{sec:leakage}

I would now like to examine a little more closely one of the assumptions made
in the above analysis.  We have ignored the possibility of {\it leakage}.  In
our model of a quantum computer, each of our qubits lives in a two-dimensional
Hilbert space.  We assumed that, when an error occurs, this qubit either
becomes entangled with the environment, or rotates in the two-dimensional space
in an unpredictable way.  But there is another possible type of error, in which
the qubit leaks out of the two-dimensional space into a larger space (Plenio \&
Knight 1996).  For example, in an ion trap computer we might store quantum
information in a two-dimensional space spanned by the ground state of the ion
and a particular long-lived metastable state (Cirac \& Zoller 1995).  But
during the operation of the device, the ion might make an unexpected transition
to another state.  If that state decays quickly to the ground state, then the
error can be detected and corrected using the now standard methods of
fault-tolerant quantum error correction.  But if the ion remains hung up in the
wrong state for a long time, those methods will fail.

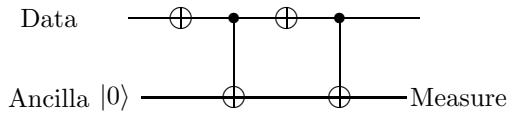
\begin{figure}
\centering
\begin{picture}(180,50)

\put(25,10){\makebox(0,0){$|0\rangle$}}
\put(0,10){\makebox(0,0){Ancilla}}
\put(0,40){\makebox(0,0){Data}}

\put(35,10){\line(1,0){100}}
\put(30,40){\line(1,0){110}}

\put(50,44){\line(0,-1){8}}
\put(50,40){\circle{8}}

\put(70,40){\circle*{4}}
\put(70,40){\line(0,-1){34}}
\put(70,10){\circle{8}}

\put(90,44){\line(0,-1){8}}
\put(90,40){\circle{8}}

\put(110,40){\circle*{4}}
\put(110,40){\line(0,-1){34}}
\put(110,10){\circle{8}}

\put(155,10){\makebox(0,0){Measure}}

\end{picture}
\caption{A quantum leak detection circuit.  The outcome of the measurement is 0
if leakage has occurred, 1 otherwise.}
\label{fig_leak}
\end{figure}

One strategy for dealing with the leakage problem would be to identify the
levels that are the most likely candidates for leakage, and to repeatedly pump
these levels back to the ground state, but this strategy might be unwieldy if
there are many candidate levels.  A more elegant solution is to detect the
leakage, but without trying to diagnose exactly what happened to the leaked
qubit.  For example, in an ion trap, we might design a controlled-NOT gate that
acts trivially if the control bit does not reside in the two-dimensional space
where the qubit belongs.  Then we can carry out the gate sequence shown in
Fig.~\ref{fig_leak}.  Nothing happens if the data qubit has leaked, but in the
event that no leakage has occurred, the ancilla bit will flip.  We then measure
the ancilla bit.  The measurement result 1 projects the qubit to the right
Hilbert space, but the result 0 projects the qubit to a leaked state.

If leakage has occurred, the qubit is damaged and must be
discarded.\footnote{Of course, we can recycle it later after it returns to the
ground state.}  We replace it with a fresh qubit in a standard state, say the
state $|0\rangle$. (If concatenated coding is used, leakage detection need be
implemented only at the lowest coding level.) Now we can perform conventional
syndrome measurement, which will project the qubit onto a state such that the
error can be reversed by a simple unitary transformation.  In fact, since we
know before the syndrome measurement that the damaged qubit is in a particular
position within the code block, we can apply a streamlined version of error
correction designed to diagnose and reverse the error at that known position
(Grassl {\it et al.} 1996).

\section{Dream machine}
\label{dream}
Let's recall some of the important assumptions that we made in our estimate of
the accuracy threshold:

\begin{itemize}
\item {\bf Random errors.} We have assumed that the errors have no systematic
component.\footnote{Knill {\it et al.} (1996, 1997) have demonstrated the
existence of an accuracy threshold for much more general error models.}  This
assumption enables us to add probabilities, rather than amplitudes, to estimate
how the probability of error accumulates over time.  Systematic errors would
accumulate much more rapidly, and therefore the error rate that can be
tolerated would be much lower.  Roughly, if the accuracy threshold is
$\epsilon_0$ for random errors, it would be of order $(\epsilon_0)^2$ for
maximally conspiratorial systematic errors.  My attitude is that (1) even if
our hardware is susceptible to making errors with systematic phases, these will
tend to cancel out in the course of a reasonably long computation (Obenland \&
Despain 1996ab; Miquel {\it et al.} 1997), and (2) since systematic errors can
in principle be understood and eliminated, from a fundamental point of view it
is more relevant to know the limitations on the performance of the machine that
are imposed by the random errors. 

\item {\bf Uncorrelated errors.} We have assumed that the errors are both
spatially and temporally uncorrelated with one another.  Thus when we say that
the probability of error per qubit is $\epsilon\sim 10^{-5}$, we actually mean
that, given two specified qubits, the probability that errors afflict both is
$\epsilon^2\sim 10^{-10}$.  This is a strong assumption, and an important one,
since our coding schemes will fail if several errors occur in the same block of
the code.  Future quantum engineers will face the challenge of ensuring that
this assumption is justified reasonably well in the devices that they design.

\item {\bf Maximal parallelism.} We have assumed that many quantum gates can be
executed in parallel in a single time step.  This assumption enables us to
perform error recovery in all of our code blocks at once, and so is critical
for controlling qubit storage errors.  (Otherwise, if we added a level of
concatenation to the code, each individual resting qubits would have to wait
longer for us to get around to perform recovery, and would be more likely to
fail.) If we ignore storage errors, then parallel operation is not essential in
the analysis of the accuracy threshold, but it would certainly be desirable to
speed up the computation.

\item {\bf Error rate independent of number of qubits.}  We have assumed that
the error rates do not depend on how many qubits are stored in our device.
Implicitly, this is an assumption about the nature of the hardware.  For
example, it would not be a reasonable assumption if all of the qubits were
stored in a single ion trap, and all shared the same phonon bus (Cirac \&
Zoller 1995).

\item {\bf Gates can act on any pair of qubits.}  We have assumed that our
machine is equipped with a set of fundamental gates that can be applied to any
pair of stored qubits (or triplet of qubits, in the case of the Toffoli gate),
irrespective of their proximity.  In practice, there is likely to be a cost,
both in processing time and error rate, of shuttling the qubits around so that
a gate can act effectively on a particular pair.  We leave it to the machine
designer to choose an architecture that minimizes this cost.  

\item {\bf Fresh ancilla qubits.} We have assumed that our computer has access to an adequate supply of fresh ancillary qubits.  The ancilla qubits are used both to implement (Toffoli) gates and to perform error recovery.  As the effects of random errors accumulate, entropy is generated, and the error recovery process flushes the entropy from the computing device into the ancilla registers.  In principle, the computation can proceed indefinitely as long as fresh ancilla qubits are provided, but in practice we will want to clear the ancilla and reuse it.  Erasing the ancilla will necessarily dissipate power and generate heat; thus cooling of the device will be required. 

\item {\bf No leakage errors.}  Leakage errors have been ignored.  But as noted
in \S~\ref{sec:leakage}, including them would not have a big effect on the
conclusions.

\end{itemize}

Our assumptions have been sufficiently realistic that we are justified in
concluding that a quantum computer with about a million qubits and an error
rate per gate of about one in a million would be a powerful and valuable device
(assuming a reasonable processing speed).  From the perspective of the current
state of the technology (Monroe {\it et 1al.} 1995; Turchette {\it et al.} 1995;
Cory {\it et al.} 1996; Gershenfeld \& Chuang 1997), these numbers seem
daunting.  But in fact a machine that meets far less demanding specifications
may still be very useful (Preskill 1997).  First of all, quantum computers can
do other things besides factoring, and some of these other tasks (in particular
quantum simulation --- Lloyd 1996) might be accomplished with a less reliable
or smaller device.  Furthermore, our estimate of the accuracy threshold might
be too conservative for a number of reasons. For example, the estimate was
obtained under the assumption that phase and amplitude errors in the qubits are
equally likely.  With a more realistic error model better representing the
error probabilities in an actual device, the error correction scheme could be
better tailored to the error model, and a higher error rate could be tolerated.
 Also, even under the assumptions stated, the fault-tolerant scheme has not
been definitively analyzed; with a more refined analysis, one can expect to
find a somewhat higher accuracy threshold, perhaps considerably higher.
Substantial improvements might also be attained by modifying the fault-tolerant
scheme, either by finding a more efficient way to implement a universal set of
fault-tolerant gates, or by finding a more efficient means of carrying out the
measurement of the error syndrome.  With various improvements, I would not be
surprised to find that a quantum computer could work effectively with a
probability of error per gate, say, of order $10^{-4}$. (That is $10^{-4}$ may
turn out to be comfortably {\it below} the accuracy threshold. In fact,
estimates of the accuracy threshold that are more optimistic than mine have
been put forward by Zalka (1996).)  

Anyway, if we accept these estimates at face value, we now have a rough target
to aim at: $10^6$ qubits with a $10^{-6}$ error rate.  That sounds pretty
tough.  But of course it might have been worse.  If we had found that we need
an error rate of order, say, $10^{-20}$, then the future prospects for quantum
computing would be dim indeed.  An error rate of $10^{-6}$ is surely ambitious,
but not, perhaps, beyond the scope of what might be achievable in the future.
In any case, we now have a fair notion of how good the performance of a useful
quantum computer will need to be.  And that in itself represents enormous
progress over just a year ago.

\bigskip

This work has been supported in part by the Department of Energy under Grant
No. DE-FG03-92-ER40701, and by DARPA under Grant No. DAAH04-96-1-0386
administered by the Army Research Office. I am grateful to David DiVincenzo and
Wojciech Zurek for organizing this stimulating meeting, and I thank Andrew Steane and Christof Zalka for helpful comments on the manuscript. I also wish to thank my
collaborators David Beckman, Jarah Evslin, Sham Kakade, and especially Daniel
Gottesman for many productive discussions about fault-tolerant quantum
computation. 

\bigskip

\leftline{\Large \bf References}

\begin{description}
\item Aharonov, D. \& Ben-Or, M. 1996a Fault tolerant quantum computation with
constant error. (Online preprint quant-ph/9611025.)
\item Beckman, D., Chari, A., Devabhaktuni, S. \& Preskill, J. 1996 Efficient
networks for quantum factoring.  {\it Phys. Rev.} A {\bf 54}, 1034.
\item Bennett, C., DiVincenzo, D., Smolin, J. \& Wootters, W. 1996
Mixed state entanglement and quantum error correction. {\it Phys. Rev.} A
{\bf 54}, 3824.
\item Calderbank, A.~R. \& Shor, P.~W. 1996 Good quantum
error-correcting codes exist. {\it Phys. Rev.} A {\bf 54}, 1098.
\item Calderbank, A.~R., Rains, E.~M., Shor, P.~W. \& Sloane, N.~J.~A. 1996
Quantum error correction via codes over GF(4). (Online preprint
quant-ph/9608006.)
\item Calderbank, A.~R., Rains, E.~M., Shor, P.~W. \& Sloane, N.~J.~A. 1997
Quantum error correction and orthogonal geometry. {\it Phys.
Rev. Lett.} {\bf 78}, 405.
\item  Cirac, J.~I. \& Zoller, P. 1995 Quantum computations
with cold trapped ions. {\it Phys. Rev. Lett.} {\bf 74}, 4091.
\item Cory, D.~G., Fahmy, A.~F. \& Havel, T.~F. 1996 Nuclear magnetic resonance
spectroscopy: an experimentally accessible paradigm for quantum computing. In
{\it Proceedings of the 4th Workshop on Physics and Computation}, Boston: New
England Complex Systems Institute.
\item Dieks, D. 1982 Communication by electron-paramagnetic-resonance devices. {\it Phys. Lett.} A {\bf 92}, 271.
\item DiVincenzo, D. \& Shor, P. 1996 Fault-tolerant error
correction with efficient quantum codes. {\it Phys. Rev. Lett.} {\bf 77},
3260.
\item Gershenfeld, N. \& Chuang, I. 1997 Bulk spin resonance
quantum computation. {\it Science} {\bf 275}, 350.
\item Griffiths, R.~B. \& Niu, C. 1996 Semiclassical Fourier transform for
quantum computation. {\it Phys. Rev. Lett.} {\bf 76}, 3228.
\item Gottesman, D. 1996 Class of quantum error-correcting
codes saturating the quantum Hamming bound. {\it Phys. Rev.} A {\bf 54}, 1862.
\item Gottesman, D. 1997a A theory of fault-tolerant
quantum computation. (Online preprint quant-ph/9702029.)
\item Gottesman, D. 1997b Stabilizer codes and quantum error correction. Ph.D.
thesis, California Institute of Technology.
\item Gottesman, D., Evslin, J. Kakade, S. \& Preskill, J. 1996, to be
published.
\item Grassl, M., Beth, Th. \& Pellizzari, T. 1996 Codes for the
quantum erasure channel. (Online preprint quant-ph/9610042.)
\item Hughes, R. 1997 (private communication).
\item Kitaev, A.~Yu. 1996a Quantum error correction with imperfect gates, preprint.
\item Kitaev, A.~Yu. 1996b Quantum computing: algorithms and error correction, preprint (in Russian).
\item Knill, E. \& Laflamme, R. 1996 Concatenated
quantum codes. (Online preprint quant-ph/9608012.)
\item Knill, E. \& Laflamme, R. 1997. A theory of
quantum error-correcting codes. {\it Phys. Rev.} A {\bf 55}, 900.
\item Knill, E., Laflamme, R. \& Zurek, W. 1996 Accuracy
threshold for quantum computation. (Online preprint quant-ph/9610011.)
\item Knill, E., Laflamme, R. \&  Zurek, W. 1997 Resilient quantum computation:
error models
and thresholds. (Online preprint quant-ph/9702058.)
\item Laflamme, R., Miquel, C., Paz, J.~P., \& Zurek, W. 1996
Perfect quantum error correction code. {\it Phys. Rev. Lett.} {\bf 77}, 198.
\item Landauer, R. 1995 Is quantum mechanics useful? {\it Phil. Tran. R. Soc.
Lond.} {\bf 353}, 367.
\item Landauer, R. 1996 The physical nature of information.  {\it Phys. Lett.}
A {\bf 217}, 188.
\item Landauer, R. 1997 Is quantum mechanically coherent computation useful? In
{\it Proc. Drexel-4 Symposium on Quantum Nonintegrability-Quantum-Classical
Correspondence}, Philadelphia, PA, 8 September 1994 (ed. D.~H. Feng and B.-L.
Hu), Boston: International Press.
\item Lenstra, A.~K., Cowie, J., Elkenbracht-Huizing, M., Furmanski, W.,
Montgomery, P.~L., Weber, D. \& Zayer, J. 1996 RSA factoring-by-web: the
world-wide status. (Online document {\tt http:
//www.npac.syr.edu/factoring/status.html}.)
\item Lloyd, S. 1996 Universal quantum simulators. {\it Science} {\bf 273},
1073.
\item Lloyd, S. 1997 The capacity of a noisy quantum channel.
{\it Phys. Rev.} A {\bf 55}, 1613.
\item MacWilliams, F.~J. \& Sloane, N.~J.~A. 1977 {\it
The Theory of Error-Correcting Codes}. New York: North-Holland Publishing
Company.
\item Miquel, C. Paz, J.~P. \& Zurek, W.~H. 1997 Quantum computation with phase
drift errors. (Online preprint quant-ph/9704003.)
\item Monroe, C., Meekhof, D.~M., King, B.~E., Itano, W.~M. \&
Wineland, D.~J. 1995 Demonstration of a fundamental quantum logic gate. {\it
Phys. 
Rev. Lett.} {\bf 75}, 4714.
\item Obenland, K. \& Despain, A.~M. 1996a Simulation of factoring on a quantum
computer architecture. In {\it Proceedings of the 4th Workshop on Physics and
Computation}, Boston, November 22-24, 1996, Boston: New England Complex Systems
Institute.
\item Obenland, K. \& Despain, A.~M. 1996b Impact of errors on a quantum
computer architecture. (Online preprint {\tt
http://www.isi.eu/acal/quantum/quantum\_op\_errors.ps}.)
\item Plenio, M.~B.  \& Knight, P.~L. 1996 Decoherence limits to quantum
computation using trapped ions. (Online preprint quant-ph/9610015.)
\item Preskill, J. 1997 Quantum computing: pro and con. (Online preprint
quant-ph/9705032.)
\item Shor, P. 1994 Algorithms for quantum computation:
discrete logarithms and factoring. In {\it Proceedings of the 35th Annual
Symposium
on Fundamentals of Computer Science}. Los Alamitos, CA: IEEE Press, pp.
124-134.
\item Shor, P. 1995 Scheme for reducing decoherence in quantum
memory. {\it Phys. Rev. A} {\bf 52}, 2493.
\item Shor, P. 1996 Fault-tolerant quantum computation. In {\it Proceedings of
the Symposium on the Foundations of Computer Science}. Los Alamitos, CA: IEEE
Press (Online preprint  quant-ph/9605011).
\item Shor, P. 1997, these proceedings.
\item Shor, P. \& and Smolin, J. 1996 Quantum error-correcting codes
need not completely reveal the error syndrome. (Online preprint
quant-ph/9604006.)
\item Steane, A.~M. 1996a Error correcting codes in quantum
theory. {\it Phys. Rev. Lett.} {\bf 77}, 793.
\item Steane, A.~M. 1996b Multiple particle interference and quantum
error correction. {\it Proc. Roy. Soc. Lond.} A {\bf 452}, 2551.
\item Steane, A.~M. 1997 Active stabilization, quantum
computation and quantum state synthesis. {\it Phys. Rev. Lett.} {\bf 78},
2252.
\item Turchette, Q.~A., Hood, C.~J., Lange, W., Mabuchi, H. \&
Kimble, H.~J. 1995 Measurement of conditional phase shifts for quantum logic. 
{\it Phys. Rev. Lett.} {\bf 75}, 4710 (1995).
\item Vedral, V., Barenco, A. \& Ekert, A. 1996 Quantum networks for
elementary arithmetic operations. {\it Phys. Rev.} A {\bf 54}, 139
\item Unruh, W.~G. 1995 Maintaining coherence in quantum computers. {\it Phys.
Rev.} A {\bf 51}, 992.
\item Wootters, W.~K. \& Zurek, W.~H. 1982 A single quantum
cannot be cloned. {\it Nature} {\bf 299}, 802.
\item Zalka, C. 1996 Threshold estimate for fault tolerant quantum
computing. (Online preprint quant-ph/9612028.)

\end{description}
\end{document}